\documentclass[amsmath,amssymb,aps,floats,amsfonts,notitlepage,superscriptaddress,eqsecnum,nofootinbib]{revtex4-1}

\selectfont

\usepackage{graphicx}       
\usepackage{dcolumn}        
\usepackage{bm}             
\usepackage{amssymb}
\usepackage{amstext}
\usepackage{tensor}
\usepackage{hyperref}
\usepackage{color}
\usepackage[usenames,dvipsnames,svgnames,table]{xcolor}
\usepackage{float}

\newcommand{\nn}{\nonumber}

\newcommand{\beq}{\begin{equation}}
\newcommand{\eeq}{\end{equation}}

\newcommand{\Lie}{\mathcal{L}}

\newcommand{\EE}{\mathcal{E}}
\newcommand{\BB}{\mathcal{B}}

\def\etal{\textit{et al.}}

\begin{document}

 \title{Tidal invariants for compact binaries on quasi-circular orbits}

\author{Sam R.~Dolan}
\email{s.dolan@sheffield.ac.uk}
\affiliation{Consortium for Fundamental Physics, School of Mathematics and Statistics,
University of Sheffield, Hicks Building, Hounsfield Road, Sheffield S3 7RH, United Kingdom.}

\author{Patrick Nolan}
\affiliation{School of Mathematical Sciences and Complex \& Adaptive Systems Laboratory, University College Dublin, Belfield, Dublin 4, Ireland.}

\author{Adrian C.~Ottewill}
\affiliation{School of Mathematical Sciences and Complex \& Adaptive Systems Laboratory, University College Dublin, Belfield, Dublin 4, Ireland.}

\author{Niels Warburton}
\affiliation{School of Mathematical Sciences and Complex \& Adaptive Systems Laboratory, University College Dublin, Belfield, Dublin 4, Ireland.}

\author{Barry Wardell}
\affiliation{Department of Astronomy, Cornell University, Ithaca, NY 14853, USA.}

\date{\today}

\begin{abstract}
We extend the gravitational self-force approach to encompass `self-interaction' tidal effects for a compact body of mass $\mu$ on a quasi-circular orbit around a black hole of mass $M \gg \mu$. Specifically, we define and calculate at $O(\mu)$ (conservative) shifts in the eigenvalues of the electric- and magnetic-type tidal tensors, and a (dissipative) shift in a scalar product between their eigenbases. This approach yields four gauge-invariant functions, from which one may construct other tidal quantities such as the curvature scalars and the speciality index. First, we analyze the general case of a geodesic in a regular perturbed vacuum spacetime admitting a helical Killing vector and a reflection symmetry. Next, we specialize to focus on circular orbits in the equatorial plane of Kerr spacetime at $O(\mu)$. We present accurate numerical results for the Schwarzschild case for orbital radii up to the light-ring, calculated via independent implementations in Lorenz and Regge-Wheeler gauges. We show that our results are consistent with  leading-order post-Newtonian expansions, and demonstrate the existence of additional structure in the strong-field regime. We anticipate that our strong-field results will inform (e.g.)~effective one-body models for the gravitational two-body problem that are invaluable in the ongoing search for gravitational waves.
\end{abstract}

\maketitle

\section{\label{sec:introduction}Introduction}
Einstein's theory of general relativity provides the framework for our modern understanding of structure formation in an expanding and accelerating cosmos. Over the century since its inception, Einstein's theory has been subjected to a battery of tests, via phenomena such as the deflection of starlight (1919), the Shapiro time delay (1966) and the precession of gyroscopes in freefall (2011). Gravitational waves (GWs) -- propagating ripples in spacetime -- are a key prediction of Einstein's theory. Strong indirect evidence for the existence of GWs comes in the form of observations of the orbital decay of the Hulse-Taylor binary in the decades since its discovery in 1974 \cite{Weisberg:2010zz}. Four decades on, in 2014, a detection of (apparently) primordial B-modes in the Cosmic Microwave Background radiation has generated much excitement, as it has been interpreted as the signature of gravitational waves in the inflationary epoch \cite{Ade:2014xna}.

The challenge of making a \emph{direct} detection of GWs from astrophysical sources is ongoing, with progress being made on two fronts. On the experimental side, a new generation of exquisitely-sensitive gravitational-wave interferometers, such as Advanced LIGO, will come online shortly. On the theoretical side, myriad improvements in models of sources \& signatures are informing strategies for data analysis.

Compact binaries featuring neutron stars and black holes are one the most promising targets for GW detectors. The challenge of modelling typical sources has led to the development of a number of complementary methodologies for attacking the gravitational two-body problem in relativity, such as post-Newtonian (PN) expansions \cite{Blanchet:2014}, gravitational self-force (GSF) theory \cite{Barack:2009, Poisson:Pound:Vega:2011}, numerical relativity (NR) and the effective-one-body (EOB) formalism \cite{Buonanno:Damour:1999, Taracchini:2014}. The first three approaches may be harnessed together to spur the fourth, as the EOB formalism provides a physically-motivated framework for synthesis. The waveforms produced by the EOB model are a crucial input for the matched-filtering approach to data analysis; hence, a concerted effort is underway to refine the EOB model \cite{Bini:Damour:2014a, Bini:Damour:2014}.

In this article, we focus on a restricted version of the gravitational two-body problem, in which two compact bodies are in a (quasi-)circular orbit. We focus on several physical quantities which can be fruitfully compared between formalisms. Specifically, we focus on the eigenvalues and eigenvectors of the electric-type and magnetic-type tidal tensors, and we isolate four independent degrees of freedom. We show that other interesting quantities, such as curvature scalars (e.g.~the Kretschmann scalar) and the speciality index, can be expressed in terms of these four. We describe a practical method for computing these quantities at $O(\mu)$ using GSF theory for equatorial orbits on Kerr spacetime, and we present a high-precision numerical calculation for the Schwarzschild case.

GSF theory seeks key results in the form of an expansion in the mass ratio $\eta = \mu / M$, where $\mu$ and $M$ are the masses of the two bodies. The mass ratio is assumed to be small $\eta \ll 1$. An appealing perspective offered by GSF theory is that the motion of the small body may be mapped onto that of a point-particle endowed with multipole moments following a trajectory in a certain regular perturbed spacetime $g^R$. Much work has been devoted to establishing this correspondence at a formal level. For example, identifying the correct regular spacetime was the focus of pioneering work in \cite{Detweiler:2001, Detweiler:Whiting:2003}.

Comparing results from GSF theory with other approaches is not necessarily straightforward, largely due to the coordinate freedom inherent in general relativity. However, a focus on computing the functional relationships between \emph{conservative gauge-invariant quantities} in GSF theory has paid dividends. Gauge-invariant quantities make up part of a Rosetta stone for translating between formalisms. \emph{Conservative} quantities cannot be computed merely from the knowledge of GW fluxes. In 2008, Detweiler \cite{Detweiler:2008} isolated the first conservative gauge-invariant relationship within GSF theory. More precisely, he studied the functional relationship between the so-called redshift invariant and the frequency of the quasi-circular orbit $\Omega$, at $O(\mu)$ for quasi-circular orbits on Schwarzschild spacetime. This led to the first successful comparison with PN theory \cite{Detweiler:2008}, and checks on GSF theory \cite{Sago:Barack:Detweiler:2008}. This comparison was shortly followed by calculations of the conservative shift at $O(\mu)$ in the innermost stable circular orbit (ISCO) \cite{Barack:Sago:2009}, and the periastron advance of eccentric orbits \cite{Barack:Sago:2011}. This strand of work led on to comparisons of PN, GSF and NR data \cite{Favata:2010, Blanchet:2010, Detweiler:LeTiec:Whiting:2010, Blanchet:2011, LeTiec:Barack:2011}, and the refinement of EOB models \cite{Damour:2010, Barack:Damour:Sago:2010, Barausse:Buonanno:LeTiec:2011, Akcay:Barack:Damour:Sago:2012}.

Recently, a second conservative gauge-invariant for circular orbits has been identified: the geodetic spin precession per unit angle, $\psi$. In Ref.~\cite{Dolan:etal:2014} the functional relationship between $\psi$ and $\Omega$ was computed at $O(\mu)$, via the standard (numerical) GSF approach, for a compact body with small spin $|\mathbf{s} | \ll G \mu^2 / c$ on a circular orbit about a large non-spinning (Schwarzschild) black hole. The precession is associated with parallel-transport in a regular perturbed spacetime; alternatively, at $O(\mu)$ it may be associated with a `self-torque' acting in the background spacetime \cite{Harte:2012}.  In Ref.~\cite{Bini:Damour:2014}, $\psi$ was calculated through $O(\mu)$ via an \emph{analytic} GSF approach, taken up to $8.5$ PN order. Impressively, analytical results were shown to capture the strong-field features of the numerical results, including the zero-crossing near the ISCO. These analytic results for $\psi$ were put to immediate use in enhancing the EOB model for spinning binaries in Ref.~\cite{Bini:Damour:2014}.

Conservative gauge-invariant quantities for circular orbits are linked to the existence of a helical Killing vector field $k^a$ in $g^R$ that coincides with the particle's tangent vector $u^a$ on the quasi-circular orbit itself. Conservative invariants may be classified according to highest derivative of $g^R$ (or equivalently $k^a$) that appears. Detweiler's redshift invariant has zero derivatives (it is formed directly from $g^R$), whereas the precession invariant features first derivatives of $g^R$. In Ref.~\cite{Bini:Damour:2014} Bini \& Damour made the argument that (i) there are no further independent invariants at zero-derivative or first-derivative order, and (ii) at second-derivative order, there are several new invariants, including the independent eigenvalues of the electric-type and magnetic-type tidal tensors. Concurrently and independently, a similar argument was put forward by Dolan \cite{Dolan:BritGrav:2014}.

In this article we describe a practical scheme for computing the shifts in these eigenvalues at $O(\mu)$ for equatorial circular orbits on Kerr spacetime, and we present highly accurate numerical results for the Schwarzschild case. The article is organised as follows. 
In Sec.~\ref{subsec:tidaltensors} we recap the theory of tidal tensors and their physical interpretation. In Sec.~\ref{subsec:invariants} we take a general approach by considering geodesic motion in a regular spacetime that admits a helical Killing vector. Here, we seek covariant expressions for tidal eigenvalues and curvature invariants. In Sec.~\ref{subsec:testparticle} we briefly describe the `test-particle' case (i.e.~the $\mu = 0$ limit). In Sec.~\ref{subsec:perturbationtheory} we apply perturbation theory to obtain formal expressions for (gauge-invariant) shifts at $O(\mu)$ in terms of the Detweiler-Whiting $R$ field. In Sec.~\ref{subsec:tidalBH} we review the theory of tidally-perturbed black holes, and extract the leading terms in the PN expansion for the eigenvalues at $O(\mu)$~\cite{Taylor:Poisson:2008, JohnsonMcDaniel:2009}. In Section \ref{sec:method} we outline the ingredients that make up frequency-domain GSF calculations in Lorenz and Regge-Wheeler gauges. In particular, in Sec.~\ref{sec:mode-sum} we provide mode-sum regularization parameters. In Sec.~\ref{sec:results} we present a selection of numerical results. We conclude in Sec.~\ref{sec:discussion} with a discussion of the implications and extensions of our work.

Throughout, we set $G = c = 1$ and use a metric signature $+2$. In certain contexts where the meaning is clear we also adopt the convention that $M = 1$. General coordinate indices are denoted with Roman letters $a,b,c, \ldots$ and indices with respect to a triad are denoted with letters $i, j, k, \ldots$. The coordinates $(t, r, \theta, \phi)$ denote general polar coordinates which, on the background Kerr spacetime, correspond to Boyer-Lindquist coordinates. Covariant derivatives are denoted using the semi-colon notation, e.g., $k_{a;b}$, with partial derivatives denoted with commas. Symmetrization and anti-symmetrization of indices is denoted with round and square brackets, $()$ and $[]$, respectively.

\section{Analysis\label{sec:analysis}}

\subsection{Tidal tensors\label{subsec:tidaltensors}}
Here we seek to characterise tidal effects measured by a geodesic observer. In general, using a timelike vector field $u^a$, one may decompose the the Riemann tensor $R_{abcd}$ into three irreducible parts \cite{Bel:1958, Cherubini:etal:2003}. In vacuum regions, where the Riemann tensor is equal to the Weyl tensor (which is self-dual), one may restrict attention to `electric-type' and `magnetic-type' tidal tensors only, defined by
\begin{subequations}
\label{eq:EEBB}
\begin{eqnarray}
\EE_{ac} &=& R_{abcd} u^b u^d ,  \label{eq:EE} \\
\BB_{ac} &=& R^*_{abcd} u^b u^d,  \label{eq:BB}
\end{eqnarray}
\end{subequations}
where $R^\ast_{abcd} \equiv \frac{1}{2} \tensor{\varepsilon}{_{ab} ^{ef}} R_{efcd}$ is the (left) Hodge dual of the Riemann tensor. Here $\epsilon_{a b c d}$ is the Levi-Civita tensor.

From the symmetries of the Riemann tensor it follows that the tidal tensors are symmetric in their indices ($\EE_{ab} = \EE_{ba}$ and $\BB_{ab} = \BB_{ba}$), and spatial ($u^a \EE_{ab} = 0 = u^a \BB_{ab}$). The magnetic-type tensor is traceless in general, and in Ricci-flat spacetimes, the electric-type tensor is also traceless, that is, ${\EE^a}_a = 0 = {\BB^a}_a$ (see e.g.~\cite{Bini:Geralico:2012, Bini:Boshkayev:Geralico:2012}).

What is the physical interpretation of the tidal tensors $\EE_{ab}$ and $\BB_{ab}$? The electric-type tensor $\EE_{ab}$, also known as the \emph{tidal field}, describes tidal gravitational accelerations, i.e., the relative acceleration of two neighbouring freely-falling particles. The magnetic-type tensor, $\BB_{ab}$, referred to as the \emph{frame-drag field} in Refs.~\cite{Nichols:etal:2011, Owen:2011}, describes tidal differential frame-dragging, that is, the difference in precession experienced by two neighbouring gyroscopes in free-fall \cite{Estabrook:Wahlquist:1964, Nichols:etal:2011}.

The electric-type tensor features in the geodesic deviation equation,
\beq
\frac{D^2 \zeta^a}{d \tau^2} = - \tensor{\EE}{^a _b} \zeta^b .
\eeq
This equation describes the acceleration of a deviation vector $\zeta^a$ which is transverse to a geodesic congruence. The magnetic-type tensor features in the Papapetrou-Pirani force on a gyroscope
$
\frac{D p^a}{d \tau} = - \tensor{\BB}{^a _b} s^b
$, where $p^a$ and $s^b$ are momentum and spin vectors, respectively.
Recent works \cite{Nichols:etal:2011, Costa:Natario:2012, Costa:Natario:Zilhao:2012} have pointed out the role of the magnetic-type tensor in generating a differential precession $\Delta \Omega_a$ for gyroscopes on neighbouring geodesics: $\Delta \Omega_a = \mathcal{B}_{ab} \zeta^b$.

\subsection{Invariants on a regular spacetime\label{subsec:invariants}}
In this section we further develop the general covariant arguments advanced in
Refs.~\cite{Dolan:etal:2014, Bini:Damour:2014}, to seek certain scalar
quantities with a natural physical interpretation. We will consider a geodesic
$\gamma$ with tangent vector $u^a$ on a regular vacuum spacetime $g_{ab}$,
subject to two simplifying assumptions. First, we assume the spacetime admits a
`helical' Killing vector field $k^a$ (with the defining property $k_{(a;b)} =
0$) which is coincident with $u^a$ on the geodesic, so that $[k^a] = u^a$. Here
we adopt the bracket notation of Ref.~\cite{Bini:Damour:2014} to indicate where
tensor fields, such as $k^a$, are evaluated on the geodesic. Second, we assume
that the spacetime and geodesic share a reflection symmetry; that is, that
there is a discrete isometry under a coordinate transformation of the form
$\theta \rightarrow \pi - \theta$. This condition is satisfied by a geodesic
lying entirely in the equatorial plane of a spacetime with an equatorial symmetry.
We may classify geometric objects as `even' or `odd' under this isometry. In
particular, scalars must be even, or zero.


\subsubsection{Zero derivatives\label{subsubsec:zeroderivs}}
If the spacetime is asymptotically flat then we may invoke the `frame of the distant stars'. The frame is defined by (asymptotic) Killing vectors; in particular, $\mathcal{T}^a \equiv \partial_t^a$ and $\mathcal{\Phi}^a \equiv \partial_\phi^a$. These enable one to define two scalars, $U = \text{lim}_{r \rightarrow \infty} k_a \mathcal{T}^a / (\mathcal{T}^b \mathcal{T}_b)$ and $\Phi = \text{lim}_{r \rightarrow \infty} k_a \mathcal{\Phi}^a / (\mathcal{\Phi}^b \mathcal{\Phi}_b)$. $U$ is (the inverse of) Detweiler's redshift invariant. The ratio of these quantities defines the orbital frequency, $\Omega \equiv \Phi / U$. The functional relationship $U(\Omega)$ was explored in Refs.~\cite{Detweiler:2008, Sago:Barack:Detweiler:2008}.

\subsubsection{First derivatives}
We begin by noting that, on the geodesic $\gamma$, $k_{a ; b}$ is a simple bivector that is orthogonal to both the tangent vector $u^a$ and an `axial' vector $\omega^a$ defined by \cite{Dolan:etal:2014}
\beq
\omega^a \equiv -\frac{1}{2} \tensor{\epsilon}{^a _{b c d}} k^b k^{c ; d}.  \label{eq:axialdef}
\eeq
That is, $[k^{b} k_{a;b}] = 0 = [\omega^{b} k_{a ; b}]$ and $[\omega^a k_a] = 0$. Now let $\omega$ denote the norm of the axial vector on the geodesic, $\omega^2 \equiv [ \omega^a \omega_a ] = \frac{1}{2} [ k_{a ;b} k^{a ; b} ]$.

To appreciate the geometrical significance of $\omega^a$, we may appeal to two natural concepts: that of parallel transport, and that of Lie transport. It is straightforward to establish that the axial vector $\omega^a$ is both parallel-transported \emph{and} Lie-transported along the geodesic, that is, $[k^a \tensor{\omega}{^b _{;a}}] = 0 = [ \Lie_k \omega^a ]$, where the Lie derivative is defined by $\Lie_{k} \omega^a = \omega^b \tensor{k}{^a _{;b}} - k^b \tensor{\omega}{^a _{;b}} = \omega^b \tensor{k}{^a _{,b}} - k^b \tensor{\omega}{^a _{,b}}$. Furthermore, $\omega^a$ is `odd', as its sign is reversed under reflection in the equatorial plane.

Let us now introduce a triad $e_{i}^a$ (where $i=1\ldots3$) on $\gamma$ whose legs are orthogonal to $u^a$ and to each other ($g_{ab} e_i^a u^b = 0$ and $g_{ab} e_i^a e_j^b = \delta_{ij}$). Let this triad be `comoving' with the geodesic, in the sense that it is Lie-transported along $k^a$, i.e.,~$\Lie_{k} e_{i}^a = 0$. Lie-transporting along a Killing field preserves inner products, and thus $\{u^a, e_{i}^a\}$ is an orthonormal basis everywhere on $\gamma$. Let us choose the second leg of the triad to be parallel to the axial vector, so that $e^a_{2} \equiv \left[ \frac{1}{\omega} \omega^a \right]$. Further, let us insist that the triad is right-handed, in the sense that $[ \epsilon^{a b c d} ] = -(4!) \, [ u^{[a} e^b_1 e^c_2 e^{d]}_3 ]$. Several useful results may be established. For example, $[k^{a ; b}] = - 2 \omega e_1^{[a} e_3^{b]}$, and thus
\beq
\frac{De^a_1}{d\tau} = + \omega e_{3}^a , \quad \quad  \frac{De^a_{2}}{d\tau} = 0 , \quad \quad \frac{De^a_3}{d\tau} = - \omega e^a_{1} ,  \label{basis-parallel}
\eeq
where $D e^a_{i} / d\tau \equiv [ k^b e^a_{i ; b} ]$, and
\beq
\left[ \tensor{k}{^a _{;c}} k^{b ; c} \right] = \omega^2 \left( e_{1}^a e_{1}^b + e_{3}^a e_{3}^b \right) .
\label{kbisym}
\eeq
Note that $e^a_1$ and $e^a_3$ are `even' and $e^a_2$ is `odd' under reflection in the equatorial plane.

We may define an alternative triad $\hat{e}_i^a$ which is parallel-transported along the geodesic, such that $[k^b \hat{e}^a_{i;b}] = 0$. This triad has legs $\hat{e}_1^a = \cos( \omega \tau ) e_1^a - \sin( \omega \tau ) e_3^a$, $\hat{e}_2^a = e_2^a$, and $\hat{e}_3^a = \sin( \omega \tau ) e_1^a + \cos( \omega \tau ) e_3^a$.
Viewed in the Lie-transported basis, the parallel-transported basis undergoes simple precession in the plane $e_1^{[a} e_3^{b]}$ at a frequency per unit proper time of $\omega$. The Lie-transported triad returns to itself after one complete orbit. Viewed from the perspective of the static observer (`distant stars') the parallel-transported basis precesses around by an angle of $2 \pi \psi$ every orbit, where
\beq
\psi = 1 - \omega / \Phi .
\eeq
The functional relationship $\psi(\Omega)$ was explored in Refs.~\cite{Dolan:etal:2014, Bini:Damour:2014}.

\subsubsection{Second derivatives and tidal tensors\label{subsubsec:secondderivs}}
Now let us consider quantities involving second derivatives of the metric. Here, the Riemann tensor will play a role, as (e.g.)~$p_{a;[bc]} = \frac{1}{2} \tensor{R}{^d _{a b c}} p_{d}$. As described in Sec.~\ref{subsec:tidaltensors}, the Riemann tensor in vacuum (i.e.~the Weyl tensor) is equivalent to electric- and magnetic-type tidal tensors defined in Eqs.~(\ref{eq:EEBB}). Let us consider the $3 \times3$ matrices formed from their basis components on $\gamma$, defined by
\beq
\EE_{ij} = [ \EE_{ab} ] e^a_i e^b_j , \quad \quad \BB_{ij} = [ \BB_{ab} ] e^a_i e^b_j .
\eeq
Alternatively, the magnetic-type matrix $\BB_{ij}$ can be written as
\beq
\BB_{ij} = \frac{1}{2} \epsilon_{jkl} R_{a b c d} u^a e^b_i e^c_k e^d_l ,
\eeq
where $\epsilon_{ijk} = \epsilon_{[ijk]}$ is the Levi-Civita symbol with $\epsilon_{123} = 1$.
$\EE_{ij}$ and $\BB_{ij}$ are symmetric and traceless $3 \times 3$ matrices. In general, each has five independent components; together they account for the ten independent components of the Weyl tensor.

Now consider the eigenvalues $\{ \lambda^E_i, \lambda^B_j \}$ and eigenvectors $\{ X^k_{(E,i)}, X^k_{(B,j)} \}$ of the tidal tensors. As the matrices are symmetric, the eigenvectors are orthogonal (or, in any degenerate case, can be chosen to be orthogonal). As the matrices are traceless, the sum of the eigenvalues is zero: $\lambda_1^E + \lambda_2^E + \lambda_3^E  = 0 = \lambda_1^B + \lambda_2^B + \lambda_3^B$. Together, the eigenvectors and eigenvalues encode ten degrees of freedom, as each orthogonal eigenbasis defines three Euler angles, and each set of eigenvalues defines two independent scalars.

Let us now consider the effect of rotating the (spatial) legs of the tetrad. The matrices transform in the usual way (i.e.~$\EE \rightarrow R \EE R^T$, with $R R^T = I$). The eigenvalues are invariant under this operation. In addition, the three Euler angles that describe the rotation that maps the `electric' eigenbasis onto the `magnetic' eigenbasis are also invariant. In other words, the scalar products of the two sets of eigenvectors are invariants. In general, then, there are seven degrees of freedom which depend only on the Weyl tensor and the tangent vector (cf.~Sec.~\ref{subsec:curvature-invariants}, below), and three more which depend also on the choice of triad. As the tangent vector has three independent components, a naive counting argument suggests there are four `intrinsic' degrees of freedom describing spacetime curvature, in general (see Sec.~\ref{subsec:curvature-invariants}).

Two key observations may be made in our case of interest: an equatorial orbit with a Killing symmetry. First,  the components of the tidal matrices are constant in the Lie-transported frame. That is, for any vector $X^a$ such that $\Lie_{k} X^a = 0$,
\beq
\frac{d}{d\tau} \left( \EE_{ab} X^a X^b \right) = 0 = \frac{d}{d\tau} \left( \BB_{ab} X^a X^b \right) .
\eeq
The proof of this statement is simple in a coordinate system which is adapted to the Killing vector, such that ${k^a}_{,b} = 0$. Then Lie transport $\Lie_k X = 0$ implies that $k^b {X^a}_{,b} = 0$ and thus
\beq
u^c \left( \EE_{ab} X^a X^b \right)_{,c} = [ k^e R_{a b c d , e} X^a k^b X^c k^d] = 0.
\eeq
The final step follows from the fact that $k^c g_{a b , c} = 0$ and partial derivatives commute. Note that the eigenvectors are Lie-dragged, not parallel-transported, along the circular orbit. Furthermore, the eigenvalues are constants along the orbit.

Second, under reflection in the equatorial plane ($\theta \rightarrow \pi - \theta$), the tidal tensors transform as $[\EE_{ab}] \rightarrow [\EE_{ab}]$ and $[\BB_{ab}] \rightarrow [- \BB_{ab}]$. It follows immediately that, e.g., $[\EE_{ab} \BB^{ab}] = 0$. Our triad transforms as $e_1^a \rightarrow e_1^a$, $e_2^a \rightarrow -e_2^a$, $e_3^a \rightarrow e_3^a$ under reflection. Therefore, many components of the matrices are zero on symmetry grounds:
\begin{eqnarray}
\EE_{12} = \EE_{32} &=& 0,  \\
\BB_{11} = \BB_{22} &=& \BB_{33} = \BB_{13} = 0 .
\end{eqnarray}
From the constraints on $\EE_{ij}$ it follows that $e_2^a = \frac{1}{\omega} \omega^a$ is an electric eigenvector, and $\lambda_2^E = \EE_{22} = \frac{1}{\omega^2} \EE_{a b} \omega^a \omega^b$ is the corresponding eigenvalue. From the constraints on $\BB_{ij}$ it follows  that one of the eigenvalues is zero and, as the matrix is traceless, the remaining eigenvalues come as a pair $(\lambda^B, -\lambda^B)$.

The axial electric eigenvalue can be rewritten in a covariant way, as follows:
\beq
\lambda_2^E = -\frac{1}{\omega^{2}} [ R_{a b c d} k^a \tensor{k}{^b _{;e}} k^c k^{d ; e} ] . \label{eq:lam2a}
\eeq

We now seek expressions for the other two electric eigenvectors, which lie in the $e_1^{[a}e_3^{b]}$ plane.
A scalar field $\kappa \equiv - k_a k^a$ may be introduced to describe the norm of the Killing vector. Note that $\kappa$ is unity on the geodesic, $[ \kappa ] = 1$. It is straightforward to verify that $k^b \tensor{k}{^a _{;b}} = -k^b \tensor{k}{_b ^{;a}} = - \frac{1}{2} \left( k^b k_b \right)^{;a} = \frac{1}{2} \kappa^{;a}$. Since the Killing vector coincides with the tangent vector, which satisfies $u^b \tensor{u}{^a _{;b}} = 0$, it follows that $[ \kappa^{;a} ] = 0$. On the other hand, the second derivatives of $\kappa$ on the geodesic are not zero, in general.

Let us consider the transport of $k_{a;b}$ along the Killing field. We note that
\begin{eqnarray}
k^c k_{a ; b c}  &=&  k^c k_{a ; c b} - R_{a c b d} k^c k^d  \nn \\
 &=& \left( k^c k_{a ; c} \right)_{;b} - \tensor{k}{^{c} _{;b}} \tensor{k}{_{a ;c}}  - R_{a c b d} k^c k^d \nn \\
 &=& \frac{1}{2} \kappa_{;a b} + k_{c ; a} \tensor{k}{^c _{;b}} - \EE_{a b} .
\end{eqnarray}
The right-hand side is symmetric in its free indices, whereas the left-hand side is antisymmetric. We thus conclude that $k^c k_{a ; b c}  = 0 $ and therefore
\begin{eqnarray}
\EE_{ab} = \frac{1}{2} \kappa_{;a b} + k_{c ; a} \tensor{k}{^c _{;b}} .  \label{eq:EEalt}
\end{eqnarray}
The last term of (\ref{eq:EEalt}), rewritten in Eq.~(\ref{kbisym}), is orthogonal to the axial vector, and so we may rewrite the eigenvalue (\ref{eq:lam2a}) in an alternative form which does not explicitly feature the Riemann tensor: $\lambda_2^E = \frac{1}{2\omega^2} [ \kappa_{;ab} \omega^a \omega^b ]$.

Now consider $\EE_{13} = [ \EE_{ab} ] e_1^a e_3^b$, which is identically zero if $e_1^a$ and $e_3^a$ are aligned with electric eigenvectors. Starting from (\ref{eq:EEalt}) it is straightforward to show that
$
\EE_{13} =  \frac{1}{2} [ \kappa_{, a b} ] e_1^a e_3^b .
$
Hence, the remaining eigenvectors correspond to the eigenvectors of a $2\times 2$ Hessian matrix $H_{ij} = \kappa_{,ab} e_i^a e_j^b$ (here $i = 1,3$).

On the Kerr background, where $u^a$ is a linear combination of two Killing vectors, this Hessian matrix is degenerate $(\det H = 0)$, and $\kappa \approx 1 + c_{11} r^2 + c_{22} (\theta - \pi / 2)^2$. We may then choose $e_1^a$ to lie in the radial direction, defining $e_1^a = n^a / \sqrt{n_a n^a}$ (where $n_a = r_{,a}$), and define $e_3^a = -\tensor{\epsilon}{^a _{b c d}} u^b e_1^c e_2^d$, noting that $\kappa_{,ab} e_3^a e_3^b = 0$ in this case. It follows from Eq.~(\ref{eq:EEalt}) that, on the background, $\lambda_3^E = \omega^2$ and so $\lambda_1^E = -\lambda_2^E - \omega^2$.
In the perturbed spacetime, this relationship no longer holds.

\subsubsection{Euler angles}
The scalar products of the eigenvectors are invariant under (spatial) rotations of the triad legs. In the general case, we expect three degrees of freedom, corresponding to the three Euler angles that specify a rotation of the electric eigenbasis onto the magnetic eigenbasis. In the case with equatorial symmetry, there is just one degree of freedom, corresponding to an angle $\chi$ in the $1$-$3$ plane. We may define
\beq
\sin \chi = \delta_{ij}  X^i_{(E,1)} X^j_{(B,3)}   \label{eq:chidef}
\eeq
where $X^i$ are the components of the eigenvectors in the orthonormal tetrad basis. Here $X^i_{(E,1)}$ is the electric eigenvector associated with the `radial' direction in the background case, and $X^j_{(B,3)}$ is the magnetic eigenvector whose corresponding eigenvalue is zero. Note that $\chi = 0$ for circular equatorial orbits on the Kerr background.

\subsubsection{Weyl scalars and curvature invariants\label{subsec:curvature-invariants}}
Although the representation in terms of tidal eigenvalues/vectors is a natural one,
there are several other equivalent invariant representations of a spacetime.
A general vacuum spacetime may be described in terms of the ten independent
components of the Riemann (or equivalently Weyl) tensor. A particularly elegant
formulation of this idea was proposed by Newman and Penrose
\cite{Newman:Penrose}. In their formalism, one defines the null tetrad
$(n^a, \ell^a, m^a, \bar{m}^a)$ consisting of two real and two complex null
vectors satisfying $n_a n^a = 0$, $\ell_a \ell^a = 0$, $n_a \ell^a = -1$,
$m_a m^a = 0$ and $m_a \bar{m}^a = 1$. The components of the Weyl tensor
in this tetrad are given by a set of five complex numbers usually referred
to as the \emph{Weyl scalars}:
\begin{subequations}
\begin{eqnarray}
\Psi_0 &=& C_{abcd} \ell^a m^b \ell^c m^d, \\
\Psi_1 &=& C_{abcd} \ell^a n^b \ell^c m^d, \\
\Psi_2 &=& C_{abcd} \ell^a m^b \bar m^c n^d, \\
\Psi_3 &=& C_{abcd} \ell^a n^b \bar m^c n^d, \\
\Psi_4 &=& C_{abcd} n^a \bar m^b n^c \bar m^d.
\end{eqnarray}
\end{subequations}

If the null tetrad is chosen such that $\ell^a$ and $n^a$ are aligned with
principal null directions of the spacetime, then $\Psi_0 = 0$ and $\Psi_4 = 0$,
respectively. A specific case of this is in Petrov type D spacetimes; if the
tetrad is chosen such that $\ell^a$ and $n^a$ are aligned with the two repeated
principal null directions of the spacetime, then the frame is called the
Kinnersley frame. In general Petrov type I spacetimes, a rotation about
the real null directions can be used to instead set $\Psi_1 = 0 = \Psi_3$,
leaving $\Psi_0$ and $\Psi_4$ non-zero. This corresponds to a gauge choice
in which the longitudinal degrees of freedom are chosen to vanish, and is
therefore referred to as the \emph{transverse frame}.

Note that the Weyl scalars are not frame-independent invariants and are not true
scalars since they do not behave appropriately under coordinate transformations.
However, the ten components may be combined to produce a total of two true
scalars and two pseudoscalars (which change sign under parity inversion
coordinate transformations). There exist several different representations of
these components in terms of complete bases of scalar polynomials of the Weyl
tensor and its dual, often referred to as \emph{scalar invariants}. A
particularly simple choice of irreducible canonical basis is given by
\begin{subequations}
\begin{eqnarray}
I_1 &=& C^{abcd} C_{abcd}, \\
I_2 &=& C^{abcd} C^{*}_{abcd}, \\
J_1 &=& C^{abcd} C_{ab}{}^{ef} C_{cdef}, \\
J_2 &=& C^{abcd} C_{ab}{}^{ef} C^{*}_{cdef}.
\end{eqnarray}
\end{subequations}
The scalar $I_1$ is commonly known as the Kretschmann scalar \cite{Henry:2000} and $I_2$ is often
referred to as the Chern-Pontryagin scalar \cite{Cherubini:etal:2003}. The
even-parity invariants $I_1$ and $J_1$ are true scalars and the odd-parity
invariants $I_2$ and $J_2$ are pseudoscalars. These four scalar invariants have
a simple representation in terms of combinations of the Weyl scalars,
\begin{align}
\label{eq:I-J-Psis}
I \equiv \frac{1}{16} (I_1 - i I_2) &= 3 \Psi_2{}^2-4 \Psi_1 \Psi_3+\Psi_0 \Psi_4,
\nonumber \\
J \equiv \frac{1}{96} (J_1 - i J_2) &= \det
  \left|
  \begin{array}{ccc}
  \Psi_4 & \Psi_3 & \Psi_2 \\
  \Psi_3 & \Psi_2 & \Psi_1 \\
  \Psi_2 & \Psi_1 & \Psi_0
  \end{array}
  \right|.
\end{align}
In a transverse frame, the four scalar invariants are then given in terms of the
two complex (pseudo)scalars $\Psi_2$ and $\Psi_0 \Psi_4$, which may be computed
from $I$ and $J$ using the characteristic polynomial $\Psi_2^3 - \tfrac14 I
\Psi_2 + \tfrac14 J = 0$ along with $\Psi_0 \Psi_4 = I - 3 \Psi_2^2$. The
solutions of the characteristic equation are most easily obtained by defining
the \emph{speciality index} \cite{Baker:Campanelli},
\beq
\mathcal{S} = 27 J^2 /
I^3,  \label{eq:speciality-index}
\eeq
which measures the deviation of the spacetime from algebraic speciality;
$\mathcal{S} = 1$ if and only if the spacetime is algebraically special and the
deviation from algebraic speciality can be measured by the scalar quantity $27
J^2 - I^3 = -\Psi_0 \Psi_4 (9 \Psi_2^2 - \Psi_0 \Psi_4)^2$. Then, it can
be shown that the appropriate root for $|\mathcal{S} - 1| < 1$ has a Taylor
series about $\mathcal{S}=1$ which is given by \cite{Beetle:2004wu}
\begin{equation}
  \Psi_2 \approx \frac{J}{I}[-3 + \tfrac43(\mathcal{S}-1) + \cdots].
\end{equation}
It is worth noting that, the case where $\Psi_0 \Psi_4 \to 0$ for
$\mathcal{S} \to 1$, the transverse frame tends to the Kinnersley frame
as $\mathcal{S} \to 1$; in this case the transverse frame is commonly referred
to as the \emph{quasi-Kinnersley} frame \cite{Nerozzi:2004wv}.

In the present context where there is a well-defined equatorial plane, the
odd-parity invariants $I_2$ and $J_2$ (and other equivalent pseudoscalar
invariants such as the Euler invariant) are zero on the equatorial plane by
symmetry considerations. Similarly, assuming the null tetrad is aligned
appropriately, then the odd-parity quantities $\Im[\Psi_0]$, $\Im[\Psi_2]$,
$\Im[\Psi_4]$, $\Re[\Psi_1]$ and $\Re[\Psi_3]$ must all be zero on the
equatorial plane. An appropriate tetrad can be chosen, for example, by requiring
that $\Re[m^a]$ is odd-parity and $\Im[m^a]$, $\ell^a$ and $n^a$ are even-parity
across the equatorial plane. For the Kerr spacetime the Kinnersley tetrad
satisfies this property, and it is reasonable to assume that a quasi-Kinnersley
transverse frame of the more general spacetime considered here will also. Then,
in this frame $\Psi_1 = 0 = \Psi_3$ and we are left with just three non-zero
Weyl scalars, $\Re[\Psi_0]$, $\Re[\Psi_2]$ and $\Re[\Psi_4]$. Furthermore,
$\Re[\Psi_0]$ and $\Re[\Psi_4]$ are not independent; a boost transformation can
be used to set $\Psi_0 = \Psi_4$ on the equatorial plane.

Then, the only two independent, non-vanishing components in the equatorial plane
are $\Re[\Psi_2]$ and  $\Re[\Psi_4]$, which transform as scalars under
reflection across the equatorial plane, but may not behave as scalars under
reflections in other directions (note, however, that they can be combined to
produce the two independent true scalars $I = 3\Psi_2^2 + \Psi_4^2$ and $J =
\Psi_2(\Psi_4^2 - \Psi_2^2)$). There are therefore at most two independent,
gauge invariant curvature degrees of freedom in the equatorial plane. These can be
physically interpreted as encoding information about the Coulomb part of the
field and one component of the gravitational radiation \cite{Szekeres:1965ux}.
Other physical quantities such as the shift in angular momentum of the spacetime
and the other component of the gravitational radiation are only available by
measurements \emph{off} the equatorial plane.

\subsubsection{Tidal eigenvalues/vectors and curvature scalars}
As described in Ref.~\cite{Dennison:2012vf}, the curvature scalars can also be expressed in
terms of the tidal eigenvalues/vectors as follows,
\begin{subequations}
\label{eq:scalar-invariants-lambda}
\begin{eqnarray}
\text{Re}(I) &=& \frac{1}{2} \sum_{i=1}^3 \left( \left(\lambda^E_i\right)^2 - \left(\lambda^B_i\right)^2 \right) \\
\text{Re}(J) &=& -\frac{1}{6} \sum_{i=1}^3 \left( \lambda_i^E \right)^3 + \frac{1}{2} \sum_{i=1}^3 \sum_{j=1}^3 \lambda^E_i (\lambda^B_j)^2 \cos^2 \left( \theta^{EB}_{ij} \right).
\end{eqnarray}
\end{subequations}
Here $\lambda_i^E$ and $\lambda_j^B$ are the electric and magnetic eigenvalues, respectively, and $\cos \left( \theta_{ij}^{EB} \right)$ are defined by the scalar products of the electric and magnetic eigenvectors. Expressions for the imaginary parts of $I$ and $J$ are also given in Ref.~\cite{Dennison:2012vf}, but recall that on the equatorial plane these are identically zero. Using the equatorial symmetry and noting that $\lambda^B_3 = 0$, $\lambda^B_1 = - \lambda^B_2 = \lambda^B$ and $\lambda_1^E + \lambda_2^E + \lambda_3^E = 0$ allows us to write
\begin{subequations}
\label{subeq:IJequatorial}
\begin{eqnarray}
\left[ I \right] &=&  \left(\lambda^E_1\right)^2 + \left(\lambda^E_2\right)^2 - \left(\lambda^B\right)^2 + \lambda_1^E \lambda_2^E \quad \equiv I_{(\lambda)},  \\
\left[ J \right] &=&
- \frac{1}{2} \lambda_3^E \left( \left( \lambda^B \right)^2 + \lambda_1^E \lambda_2^E  \right) + \left[ \frac{1}{2} \lambda_3^E \left(\lambda^B\right)^2  + \frac{1}{2} \sum_{i=1}^3 \sum_{j=1}^2 \lambda^E_i (\lambda^B)^2 \cos^2 \left( \theta^{EB}_{ij} \right) \right]  \\
&\equiv& \quad \quad \quad  \quad J_{(\lambda)} \quad  \quad \quad + \quad  \quad \quad J_{(\chi)} .
\end{eqnarray}
\end{subequations}
The key advantage of splitting $[J]$ in this manner is that we may make use of the factorization
\beq
27 J^2_{(\lambda)} - I_{(\lambda)}^3 = -\frac{1}{4} \left(\lambda^E_1 - \lambda^E_2 + 2 \lambda^B  \right)   \left(\lambda^E_1 - \lambda^E_2 - 2 \lambda^B  \right) \left( \left(\lambda_B\right)^2 + 2 (\lambda_3^E)^2 + \lambda_1^E \lambda_2^E \right)^2
\label{eq:IJfactorize}
\eeq
In the equatorial Kerr case $J_{(\chi)} = O(\mu^2)$, and the repeated root in Eq.~(\ref{eq:IJfactorize}) ensures that $\mathcal{S} = 1 + O(\mu^2)$. We will show in Sec.~\ref{subsec:perturbationtheory} that the shift in $\mathcal{S}$ at $O(\mu^2)$ may be computed from quadratic combinations of $O(\mu)$ quantities.

\subsection{Circular orbits of test particles\label{subsec:testparticle}}
In this section we consider a circular geodesic orbit of a test particle ($\mu = 0$) at radius $r=r_0$ in the equatorial plane of Kerr spacetime. We make explicit the various expressions derived in the previous sections, working with Boyer-Lindquist coordinates $\{t,r,\theta,\phi\}$.

The helical Killing field $k^a$ and Lie-transported triad $e_i^a$ on $\gamma$ have the components \cite{Marck:1983}
\begin{subequations}
\begin{eqnarray}
k^a &=& [U , 0, 0, \Omega U] \\
e_1^a &=& [0, \sqrt{\Delta_0} / r_0, 0, 0]  \label{rhodef} \\
e_2^a &=& [0,0,1/r_0,0] \\
e_3^a &=& -\tensor{\epsilon}{^a _{b c d}} u^b e_1^c e_2^d
\end{eqnarray}
\end{subequations}
where $u^a = [k^a]$, $\Omega = \sqrt{M}/(r_0^{3/2} + a\sqrt{M})$, $U = \sqrt{M} / (\Omega r_0^{3/2} \upsilon)$, $\Delta_0 = r_0^2 - 2Mr_0 + a^2$ and
\beq
\upsilon^2 \equiv 1 - 3M/r_0 + 2 a \sqrt{M} / r_0^{3/2}.   \label{eq:upsilon}
\eeq
The norm of the axial vector $\omega^a$, introduced in Eq.~(\ref{eq:axialdef}), is
\beq
\omega = \frac{M^{1/2}}{r_0^{3/2}}.
\eeq
The spin precession invariant \cite{Dolan:etal:2014} is
\beq
\psi = 1 - \upsilon .
\eeq

In this basis, the only non-zero elements of the tidal matrices are $\EE_{11}$, $\EE_{22}$, $\EE_{33}$ and $\BB_{12} = \BB_{21}$. The electric-type eigenvalues are
\begin{eqnarray}
\lambda^E_1 &=& \EE_{11} = \frac{M}{r_0^3} - \frac{3 M \Delta_0}{\upsilon^2 r_0^5}, \\
\lambda^E_2 &=& \EE_{22} = -\frac{2M}{r_0^3} + \frac{3 M \Delta_0}{\upsilon^2 r_0^5}, \\
\lambda^E_3 &=& \EE_{33} = \frac{M}{r_0^3} .
\end{eqnarray}
Note that the sum of eigenvalues is zero, as expected.  Negative eigenvalues indicate tidal `stretching' (e.g.~in the radial direction) and positive values indicate tidal `compression'.

As argued in Sec.~\ref{subsubsec:secondderivs}, one of the eigenvalues of the magnetic-type tidal matrix is zero, due to equatorial symmetry, with corresponding eigenvector $e_3^a$. There remains a pair of eigenvalues $\pm \lambda^B$ and eigenvectors $\frac{1}{\sqrt{2}} \left(e_1^a \pm e_2^a \right)$, where
\beq
\lambda^B = \BB_{12} = -\frac{3 M^{3/2} \sqrt{\Delta_0}  \left(1 - a/\sqrt{M r_0} \right)}{r_0^{9/2} \upsilon^2} .
\eeq
As with all type D spacetimes, an appropriate null frame can be chosen such that
the only non-zero Weyl scalar is $\Psi_2$. In the Kerr spacetime this frame
is the Kinnersley frame and $\Psi_2$ is given by the simple expression
\begin{equation}
\Psi_2 = -\frac{M}{(r-i a \cos \theta)^3}.
\end{equation}
This simplifies further in the equatorial plane; the dependence on $a$ drops
out and $\Psi_2$ is purely real. In that case, there is just a single independent
non-zero scalar invariant given by $[\Psi_2] = -M/r_0^3$, with $[I] = 3 [\Psi_2]^2$ and $[J] = -[\Psi_2]^3$.
Using $\mathcal{S} = \tfrac{27 J^2}{I^3} = 1$ for
an unperturbed type D spacetime, we obtain an identity for the
magnetic eigenvalue,
\begin{equation}
\label{eq:lambdaB-constraint}
(\lambda^B)^2 = -2 (\lambda_3^E)^2 - \lambda^E_1 \lambda^E_2.
\end{equation}
Note that Eq.~(\ref{eq:lambdaB-constraint}) follows from the repeated factor in Eq.~(\ref{eq:IJfactorize}) (N.B.~ $J_{(\chi)} = 0 $ in the $\mu = 0$ case). Along with $\lambda^E_1+\lambda^E_2+\lambda^E_3=0$ we can therefore solve for the eigenvalues to get
\begin{equation}
  \lambda^E_1 + \lambda^E_2 = \Psi_2, \quad \lambda^E_3 = -\Psi_2, \quad
  (\lambda^B)^2 + \lambda^E_1 \lambda^E_2 = - 2 \Psi_2^2.
\end{equation}
Notice that there are now only two independent eigenvalues, $\lambda_2$ and
$\lambda_3$, in the $\mu = 0$ case. This is not the case for $\mu \neq 0$. 


\subsection{Perturbation theory\label{subsec:perturbationtheory}}
In this section, we seek expressions for the eigenvalues of the tidal matrices in the regular perturbed spacetime $\bar{g}_{ab} + h_{ab}^R$, where $\bar{g}_{ab}$ is the Kerr metric (in Boyer-Lindquist coordinates) and $h_{ab}^R = O(\mu)$ is the `regular' metric perturbation defined by Detweiler \& Whiting \cite{Detweiler:Whiting:2003}. Here we will work to first order in the small mass $\mu$, neglecting all terms at $O(\mu^2)$. Note that the regular perturbed spacetime is vacuum (i.e.~Ricci-flat).

We take a two-step approach. First, we compare quantities in the perturbed spacetime with quantities on the background spacetime \emph{which are defined at the same coordinate radius $r=r_0$}. 
Then, noting that $r_0$ itself varies under a gauge transformation at $O(\mu)$, we re-express key quantities in terms of the orbital frequency $\Omega$ (an observable) to obtain gauge-invariant functional relationships (e.g.~$\lambda_i^E(\Omega)$). Broadly speaking, this is the approach developed in Refs.~\cite{Detweiler:2008, Sago:Barack:Detweiler:2008}.

Henceforth, we will use an `over-bar' to denote quantities which take the same coordinate values as corresponding quantities on the background spacetime $\bar{g}_{ab}$. That is, barred quantities such as $\bar{u}^a$ are assigned the same coordinate values as in Sec.~\ref{subsec:testparticle}. We use $\delta$ to denote the difference at $O(\mu)$, i.e.,~$\delta e^a_i \equiv e^a_i - \bar{e}^a_i$. In general, such differences are gauge-dependent. At $O(\mu)$, $\delta$ may be applied as an operator with a Leibniz rule $\delta (AB) = (\delta A)B + A \delta B$.

To split a physical quantity, $Y$ say, into $O(\mu^0)$ and $O(\mu^1)$ parts in a well-defined way, we follow the standard GSF convention \cite{Detweiler:2008, Sago:Barack:Detweiler:2008}. First, we introduce the `frequency-radius' $r_{\Omega}$, defined via
\beq
\Omega = \sqrt{M} / (r_\Omega^{3/2} + a \sqrt{M}) . \label{eq:freqradius}
\eeq
Then, we write
\beq
Y - \bar{Y}(r_{\Omega}) = \Delta Y(r_0) + O(\mu^2).  \label{eq:Deltadef}
\eeq
Here $\bar{Y}(r_\Omega)$ has the same functional form as $Y$ on the background spacetime, but with $r_0$ replaced by $r_{\Omega}$. As $\Delta Y$ is at $O(\mu)$, we may parameterize $\Delta Y$ using the $O(\mu^0)$ `background' radius $r_0$, rather than $r_{\Omega}$, as $r_0 - r_{\Omega} = O(\mu)$ and so corrections are at higher order, $O(\mu^2)$. 

To simplify the analysis, let us work within a class of gauges in which the metric perturbation is helically-symmetric. This implies that $\bar{u}^c h^R_{a b , c} = 0$ at the relevant order.

\subsubsection{Tidal eigenvalues}

The simple form of the tidal matrices on the background spacetime in the Lie-transported basis means that it is simple to find the variation of the eigenvalues at leading order in $\mu$. We have $\delta \lambda_i^E = \delta \EE_{ii}$ (no summation) and $\delta \lambda^B = \delta \BB_{21}$, hence,
\begin{eqnarray}
\delta \lambda_i^E &=&  \delta R_{abcd} \, \bar{e}_i^a \bar{u}^b \bar{e}_i^c \bar{u}^d + \bar{R}_{abcd} \delta \left( e_i^a u^b e_i^c u^d \right) \\
\delta \lambda^B &=& \delta R_{abcd} \, \bar{u}^a \bar{e}_2^b \bar{e}_2^c \bar{e}_3^d + \bar{R}_{abcd} \delta \left(  u^a e_2^b e_2^c e_3^d \right)
\end{eqnarray}

The variation of the Riemann tensor can be found in the standard way from the metric perturbation.
The variation of the tangent vector may be found by recalling key relations
previously established in GSF theory for equatorial circular orbits on Kerr
spacetime \cite{Detweiler:2008, Shah:Friedman:Keidl:2012}, namely,
\begin{eqnarray}
\frac{\delta u^t}{\bar{u}^t} &=& \frac{1}{2} h_{00} - \frac{\bar{\Omega}}{2} \sqrt{\frac{r_0}{M}} \left( r_0^2 + a^2 - 2 a\sqrt{Mr_0} \right) \tilde{F}_r ,   \label{eq:ut} \\
\frac{\delta u^\phi}{\bar{u}^\phi} &=&  \frac{1}{2} h_{00} - \frac{1}{2M} \left( r_0^2 - 2M r_0 + a \sqrt{M r_0} \right) \tilde{F}_r  .\label{eq:uphi}
\end{eqnarray}
Here $h_{00} \equiv h^R_{a b} \bar{u}^a \bar{u}^b$, and the radial component of the GSF is given by
\beq
\tilde{F}_r \equiv \mu^{-1} F_r =  \left. \frac{1}{2} \bar{u}^a \bar{u}^b \partial_r h_{a b} \right|_{r=r_0}.
\eeq
We may use Eq.~(\ref{eq:ut}) and (\ref{eq:uphi}) to write the variation in the tangent vector as follows:
\beq
\delta u^a = \frac{1}{2} h_{00} \bar{u}^a + \beta_{03} \bar{e}_3^a ,
\quad \text{where} \quad \beta_{03} = -\frac{1}{2} \sqrt{\frac{r_0 \Delta_0}{M}} \, \tilde{F}_r.
\eeq
The legs of the triad can be expanded in a similar way, using $\delta e_i^a = \beta_{i 0} \bar{u}^a + \sum_j \beta_{ij} \bar{e}_j^a$.  The diagonal coefficients $\beta_{ii}$ are found by imposing the normalization condition, $\left(\bar{g}_{ab} + h^R_{ab}\right) \left(\bar{e}_i^a + \delta e_i^a\right) \left(\bar{e}_j^b + \delta e_j^b\right) = \delta_{ij}$. From normality, we infer that $\beta_{00} = \frac{1}{2}h_{00}$ (as above), and $\beta_{ii} = -\frac{1}{2} h_{ii}$, where $h_{ii} = h^R_{ab} \bar{e}^a_i \bar{e}^b_i$. From orthogonality of legs $1$ and $3$, we obtain $\beta_{30} = \beta_{03} + h_{03}$. It turns out that $\beta_{03}$ and $\beta_{30}$ are the only off-diagonal coefficients needed in our analysis, due to the very simple form of the background Riemann tensor in our chosen basis.

The variation in the eigenvalues may be expressed succinctly as follows:
\begin{eqnarray}
\delta \lambda_1^{E} &=& (\delta R)_{\bar{1}\bar{0}\bar{1}\bar{0}} + \left(h_{00} - h_{11} \right) \bar{\lambda}_1^E + 2 \beta_{03} \bar{\lambda}^B , \label{dellam1} \\
\delta \lambda_2^{E} &=& (\delta R)_{\bar{2}\bar{0}\bar{2}\bar{0}} + \left(h_{00} - h_{22} \right) \bar{\lambda}_2^E - 2 \beta_{03} \bar{\lambda}^B , \\
\delta \lambda_3^{E} &=& (\delta R)_{\bar{3}\bar{0}\bar{3}\bar{0}} + \left(h_{00} - h_{33} \right) \bar{\lambda}_3^E , \\
\delta \lambda^B &=& (\delta R)_{\bar{0}\bar{2}\bar{2}\bar{3}} + \frac{1}{2} \left(h_{00} - 2h_{22} - h_{33}\right) \bar{\lambda}^B + \beta_{03} \left( \bar{\lambda}_1^E - \bar{\lambda}_2^E \right) - h_{03} \bar{\lambda}_2^E ,  \label{dellamB}
\end{eqnarray}
where $h_{ij} = h^R_{ab} \bar{e}_i^a \bar{e}_j^b$, $h_{0i} = h^R_{ab} \bar{u}^a \bar{e}_i^b$ and
\begin{eqnarray}
(\delta R)_{\bar{i}\bar{0}\bar{j}\bar{0}} &=& \delta R_{a b c d} \bar{e}_i^a \bar{u}^b \bar{e}_j^c \bar{u}^d , \\
(\delta R)_{\bar{0}\bar{2}\bar{2}\bar{3}} &=& \delta R_{a b c d} \bar{u}^a \bar{e}_2^b \bar{e}_2^c \bar{e}_3^d .
\end{eqnarray}
As noted above, the coordinate radius of the orbit, $r=r_0$, is not invariant under changes of gauge (i.e.~coordinate changes at $O(\mu)$). However, recall that the orbital frequency $\Omega$ has a gauge-invariant definition, given in Sec.~\ref{subsubsec:zeroderivs}. Following Eq.~(\ref{eq:Deltadef}), we may  express the functional relationship between $\lambda$ and $\Omega$ as follows,
\beq
\lambda(\Omega) = \bar{\lambda}(r_{\Omega}) + \Delta \lambda(r_0) + O(\mu^2),
\eeq
where $r_{\Omega}$ is the frequency-radius defined in Eq.~(\ref{eq:freqradius}), $\lambda \in \{ \lambda^E_i, \lambda^B \}$ and $\Delta \lambda = O(\mu)$. Note that $\bar{\lambda}(r_\Omega)$ denote the `test-particle' functions defined in Sec.~\ref{subsec:testparticle} evaluated at $r_\Omega$. It is straightforward to show that, at $O(\mu)$,
\beq
\Delta \lambda = \delta \lambda - \delta \Omega \frac{d r_0}{d \bar{\Omega}} \frac{d \bar{\lambda}}{d r_0} ,
\eeq
or, making use of Eq.~(\ref{eq:ut}) and (\ref{eq:uphi}) and $\delta \Omega / \bar{\Omega} = \delta u^\phi / \bar{u}^\phi - \delta u^t / \bar{u}^t$,
\beq
\Delta \lambda = \delta \lambda - \frac{1}{3M} r_0^3 \upsilon^2 \tilde{F}_r \frac{d\bar{\lambda}}{dr_0} . \label{dlam}
\eeq
In summary, $\Delta \lambda$ defined by Eq.~(\ref{dlam}) and Eq.~(\ref{dellam1})--(\ref{dellamB}) are the gauge-invariant quantities we have been seeking, and which we will compute in the next section.

With the aid of a symbolic algebra package, it is straightforward to verify explicitly that $\Delta \lambda$ are invariant under any change of gauge which respects the helical symmetry. That is, under the transformation $h_{ab} \rightarrow h_{ab} - 2 \xi_{(a ; b)}$, where $u^b \tensor{\xi}{^a _{,b}} = 0$ and $[\xi^\theta] = 0$.
Furthermore, it is straightforward to verify that the traceless condition also
holds at $O(\mu)$,
\begin{equation}
\label{eq:traceless-perturbed-eigenvalues}
\Delta \lambda^E_1 + \Delta \lambda^E_2 + \Delta \lambda^E_3 = 0, 
\end{equation}
as expected from the fact that the regularized perturbed spacetime is also
vacuum. In contrast, there is no constraint equivalent to
Eq.~\eqref{eq:lambdaB-constraint} for $\Delta \lambda^B$ at $O(\mu)$.

\subsubsection{Scalar product of eigenvectors}
As discussed in previous sections, the scalar products formed between the electric- and magnetic eigenbases are well-defined quantities which do not depend on the choice of triad. In the equatorial case there is a single degree of freedom $\chi$, defined in Eq.~\ref{eq:chidef}, which is zero on the background ($\bar{\chi} = 0$). At $O(\mu)$, it is sufficient to use $\chi = \mathcal{E}_{13} / (\bar{\lambda}_1 - \bar{\lambda}_3) - \mathcal{B}_{23} / \bar{\lambda}_B$. Following the steps in the previous sections, we find
\beq
\Delta \chi = \delta \chi = \frac{(\delta R)_{\bar{0}\bar{2}\bar{2}\bar{1}} - h_{01} \bar{\lambda}_2^E}{\bar{\lambda}^B} + \frac{(\delta R)_{\bar{0}\bar{1}\bar{0}\bar{3}} - h_{13} \bar{\lambda}_1^E}{\bar{\lambda}_1 - \bar{\lambda}_3} .  \label{eq:dchi}
\eeq
Note that $\Delta \chi$ is dissipative, rather than conservative, in character,
and that it requires no regularization.

\subsubsection{Curvature scalars}

The $O(\mu)$ shift in the curvature scalars is given by
\begin{subequations}
\begin{eqnarray}
  \delta I &=& -\tfrac12 h \bar{I} - \tfrac14 (C^{abcd} - i C^{*abcd}) h_{ac;bd}, \\
  \delta J &=& -\tfrac34 h \bar{J} - \tfrac{1}{16} (C^{ab}{}_{ef}C^{cdef} - i C^{ab}{}_{ef}C^{*cdef}) h_{ac;bd}.
\end{eqnarray}
\end{subequations}
This may be given in terms of the $O(\mu)$ shift in the tidal tensors by
\begin{subequations}
\begin{eqnarray}
\,[\Delta I] &=& \,\left[\bar{\lambda}_1^{E} \Delta\lambda_1^{E} + \bar{\lambda}_2^{E} \Delta\lambda_2^{E}
                    + \bar{\lambda}_3^{E} \Delta\lambda_3^{E}- 2 \bar{\lambda}^{B} \Delta\lambda^B \right] + O(\mu^2), \label{eq:DeltaI} \\
\,[\Delta J] &=& - \tfrac12 \,\left[\bar{\lambda}_2 \bar{\lambda}_3 \Delta \lambda_1
                      + \bar{\lambda}_1 \bar{\lambda}_3 \Delta \lambda_2
                      + \left(\bar{\lambda}_1 \bar{\lambda}_2 + (\lambda^B)^2\right) \Delta \lambda_3
                      + 2 \bar{\lambda}_3 \bar{\lambda}^B \Delta \lambda^B
                    \right] + O(\mu^2).
\end{eqnarray}
\end{subequations}
We note that, due to the algebraic speciality of the background, $\mathcal{S} = 1 + O(\mu^2)$, it follows that
at $O(\mu)$,
\beq\label{eq:dI_over_I}
\frac{1}{2} \frac{\Delta I}{\bar{I}} = \frac{1}{3} \frac{\Delta J}{\bar{J}} = \frac{ \Delta \Psi_2}{ \bar{\Psi}_2}
\eeq
on the geodesic (cf.~Fig.~\ref{fig:curvature}). In the final equality, we have
assumed a quasi-Kinnersley frame where only $\Psi_2$ is non-zero in the
background.

\subsubsection{Speciality index $\mathcal{S}$}
To compute the speciality index $\mathcal{S}$ through $O(\mu^2)$ using Eq.~(\ref{subeq:IJequatorial}) we also require the square of the scalar products at $O(\mu^2)$; the relevant quantities are
\begin{subequations}
\begin{eqnarray}
\cos^2 \left( \theta^{EB}_{1j} \right) &=&  \frac{1}{2} \left(1 - (\Delta \chi)^2 \right) \\
\cos^2 \left( \theta^{EB}_{2j} \right) &=&  \frac{1}{2}  \\
\cos^2 \left( \theta^{EB}_{3j} \right) &=&  \frac{1}{2} \left( \Delta \chi \right)^2
\end{eqnarray}
\end{subequations}
where $j = 1,2$. Referring now to Eq.~(\ref{subeq:IJequatorial}), it follows that, at $O(\mu^2)$,
\beq
J_{(\chi)} = \frac{1}{2} \left( \bar{\lambda}_3 - \bar{\lambda}_1 \right) (\bar{\lambda}^B)^2  \Delta \chi^2 .
\eeq
Now, using $\Delta \mathcal{S} = \Delta \left( 27 J^2 - I^3 \right) / \bar{I}^3$, and noting the factorization (\ref{eq:IJfactorize}) with the repeated root, we obtain
\beq
\Delta \mathcal{S} = - 3 \left( \frac{\Delta \Lambda}{\bar{I}} \right)^2 + \frac{2 J_{(\chi)}}{\bar{J}} .
\label{eq:DeltaS}
\eeq
at $O(\mu^2)$ where
\begin{eqnarray}
\Delta \Lambda &\equiv& \frac{1}{2} \Delta \left[ (\lambda^B)^2 + 2 (\lambda_3^E)^2 + \lambda_1^E \lambda_2^E   \right]  \\
 &=& \bar{\lambda}^B \Delta \lambda^B + 2 \bar{\lambda}_3^E \Delta \lambda_3^E + \frac{1}{2} \left( \bar{\lambda}_1^E \Delta \lambda_2^E + \bar{\lambda}_2^E \Delta \lambda_1^E \right) .
\end{eqnarray}
Note that $\Delta \mathcal{S}$ is at order $O(\mu^2)$, but is constructed from quadratic combinations of $O(\mu)$ quantities, due to the algebraic speciality of the background. Note also that the first term in Eq.~(\ref{eq:DeltaS}) is built from the eigenvalues, which are conservative in character, whereas the second term is built from $\Delta \chi$, which is dissipative in character.

We may arrive at a similar result in terms of the Weyl scalars. Using
Eq.~\eqref{eq:I-J-Psis} in $\mathcal{S} = 27 J^2/I^3$ and expanding to
$O(\mu^2)$, we get
\begin{equation}
  \Delta \mathcal{S} = - \frac{3 \Delta \Psi_0 \Delta \Psi_4}{\bar{\Psi}_2^2},
\end{equation}
assuming a quasi-Kinnersley frame where only $\Psi_2$ is non-zero
in the background. As with the tidal invariants, we see that
$\Delta \mathcal{S}$ is $O(\mu^2)$ but is constructed from the quadratic
combination, $\Delta \Psi_0 \Delta \Psi_4$, of two first-order quantities.
Note that the form of this expression is frame dependent. Regardless of the
frame, however, it is always possible to compute $\Delta \mathcal{S}$ from
$O(\mu)$ quantities alone.

\subsubsection{Spin precession scalar}
Let us now consider the shift $\Delta \psi$ in the spin precession invariant $\psi = \bar{\psi}(r_\Omega) + \Delta \psi$ at $O(\mu)$. For the Schwarzschild case, $\Delta \psi$ is given by the right-hand side of Eq.~(7) in Ref.~\cite{Dolan:etal:2014}. Here we present an alternative analysis which leads to an expression for the equatorial Kerr case. Our starting point is an expression for the magnitude of the axial vector in terms of the Lie-transported tetrad,
\beq
\omega = \Gamma_{a b c} \, e_3^a u^b e_1^c .
\eeq
where $\Gamma_{a b c} = \frac{1}{2} \left(  g_{a c , b} + g_{a b , c} - g_{b c , a} \right)$ is the affine connection.
Applying the variation operator leads to
\beq
\delta \omega = \frac{1}{2} \left( h_{00} - h_{11} - h_{33} \right) \bar{\omega} + (\delta \Gamma)_{\bar{3}\bar{0}\bar{1}} + \beta_{03} \bar{ \Gamma}_{331}
\eeq
where $(\delta \Gamma)_{\bar{3}\bar{0}\bar{1}} = \frac{1}{2} \left( h_{a c , b} + h_{a b , c} - h_{b c , a} \right) \bar{e}_3^a \bar{u}^b \bar{e}_1^c$ and $\bar{\Gamma}_{331} = \bar{\Gamma}_{abc} \bar{e}^a_3 \bar{e}^b_3 \bar{e}^c_1$. The variation in the precession invariant is given by $\delta \psi = -\upsilon \left( \frac{\delta \omega}{\bar{\omega}} - \frac{\delta u^{\phi}}{\bar{u}^\phi} \right)$ (with $\upsilon$ defined in Eq.~(\ref{eq:upsilon})) or explicitly,
\beq
\delta \psi = \upsilon \left( - \frac{1}{\bar{\omega}} (\delta \Gamma)_{301}  + \frac{1}{2} \left(h_{11} + h_{33} \right) + \frac{1}{2} \left(r_0 - a \sqrt{r_0/M} \right) \tilde{F}_r \right)
\eeq
As before, a gauge-invariant quantity at $O(\mu)$ may be constructed by
introducing the frequency-radius, and writing $\psi = \bar\psi(r_\Omega) +
\Delta \psi$. This yields
\beq
\Delta \psi = \upsilon \left( - r_0^{3/2} M^{-1/2} (\delta \Gamma)_{301}  + \frac{1}{2} \left(h_{11} + h_{33} \right) +  \left(r_0 - a \sqrt{r_0/M} \right) \tilde{F}_r \right) . \label{eq:Dpsi}
\eeq
It is straightforward to check that, in the Schwarzschild case ($a = 0$), Eq.~(\ref{eq:Dpsi}) is equivalent to Eq.~(7) in Ref.~\cite{Dolan:etal:2014}.

\subsection{Interpretation of tidal effects\label{subsec:tidalBH}}

In this section we seek to clarify the relationship between the shifts in tidal eigenvalues, which are defined on a (fictitious) regular perturbed vacuum spacetime $\bar{g} + h^R$, and physical tidal effects, which could (in principle) be detected in the vicinity of a black hole in a binary system. Here, we may draw upon a line of work, initiated by Manasse \cite{Manasse:1963} and developed by many others \cite{DEath:1975, Kates:1980, Thorne:Hartle:1985, Alvi:2000, Price:Whelan:2001, Detweiler:2001, Detweiler:2005,
Poisson:2003, Poisson:2005, Taylor:Poisson:2008, JohnsonMcDaniel:2009}, which address a key question: how does a black hole move through, and respond to, an external environment?

The standard tool for analyzing this kind of problem is the method of matched asymptotic expansions (MAE). In essence, the existence of two very different characteristic length scales in the problem ($M \gg \mu$) allows one to construct complementary expansions in `inner' ($r \sim \mu$) and `outer' ($r \sim M$) zones that, with some delicacy, may be connected in a suitable `buffer' zone $\mu \ll r \ll M$. Indeed, the first derivation of the GSF equations of motion \cite{Mino:Sasaki:Tanaka:1997} was constructed using matched asymptotic expansions. The works of Hartle \& Thorne \cite{Thorne:Hartle:1985}, Alvi \cite{Alvi:2000}, Detweiler \cite{Detweiler:2001}, Poisson \cite{Poisson:2005}, and Yunes \emph{et al.}~\cite{Yunes:2006} also employ the method.

Although the underlying idea is straightforward, the application of matched asymptotic expansions in general relativity is greatly complicated by coordinate freedom.  As noted by Pound \cite{Pound:2010}, typically `inner' and `outer' expansions represent two different spacetimes expressed in two different coordinate systems. The existence of overlapping terms in dual expansions in a buffer region does not guarantee the existence of a well-behaved coordinate transformation between the two systems. Constructing a truly rigorous argument requires much attention to detail which is beyond the scope of this work. Here, the aim is to sketch a heuristic argument, closely modelled on the physically-motivated work of Detweiler in Refs.~\cite{Detweiler:2001} and \cite{Detweiler:2005}.

\subsubsection{Tidally-perturbed black holes}
Let us first consider the `outer' expansion. Through $O(\mu)$, the work of Detweiler \& Whiting \cite{Detweiler:Whiting:2003} has established that motion of a `small' non-rotating black hole is associated with a geodesic worldline $\gamma$ in a regular perturbed spacetime $g_{ab}^R = \bar{g}_{ab} + h_{ab}^R$. We may introduce a parallel-transported tetrad $\{ u^a, \hat{e}_i^a \}$ on $\gamma$, noting that the parallel-transported tetrad is distinct from the Lie-transported basis of Sec.~\ref{subsec:invariants}. Using this tetrad, we may construct a Fermi normal coordinate system in vicinity of the worldline, on which $[g_{ab}^R] = \eta_{ab}$ and $[\tensor{\Gamma}{^a _{bc}}] = 0$. A further coordinate transformation takes us to Thorne-Hartle-Zhang (THZ) coordinates \cite{Thorne:Hartle:1985, Zhang:1986} $\{ \hat{t}, \hat{x}^i \}$, $\hat{r} = \sqrt{\hat{x}^2 + \hat{y}^2 + \hat{z}^2}$ in which the metric takes the form
\beq
g^R_{ab} = \eta_{ab} + {}_2H_{ab} + O(\hat{r}^3 / M^3) ,
\eeq
where
\beq
{}_2H_{ab} d\hat{x}^a d\hat{x}^b = -\hat{\mathcal{E}}_{ij} \hat{x}^i \hat{x}^j \left( d\hat{t}^2 + \delta_{kl} d\hat{x}^k d\hat{x}^l \right) + \frac{4}{3} \epsilon_{k p q} \tensor{ \hat{\mathcal{B}} }{^q _i} \hat{x}^p \hat{x}^i d \hat{t} d \hat{x}^k + \mathcal{O}(\hat{r}^3 / M^3) \label{2Hab} .
\eeq
In the vicinity of the worldline the metric looks locally flat, but with a quadrupolar term encoding tidal effects.
Here $\hat{\mathcal{E}}_{ij}$ and $\hat{\mathcal{B}}_{ij}$ are formed by projecting the Riemann tensor of the regular perturbed spacetime (and its dual) onto the parallel-transported basis. Note that we have neglected terms in (\ref{2Hab}) involving time derivatives of the tidal tensors, which, though non-zero due to the precession of the (Lie-dragged) body frame relative to the parallel-transported spin frame, are suppressed by an additional factor of $\hat{r} / M$. At the next order in $\hat{r} / M$, the expansion also features octupolar terms. For a more complete analysis, see Sec.~3 in Ref.~\cite{Detweiler:2005}.

For the `inner' solution, we may start with the metric for a tidally-perturbed Schwarzschild black hole,
\beq
g_{ab} = g_{ab}^{\text{Schw}}(\mu) + {}_2 h_{ab} + \ldots
\eeq
where $g_{ab}^{\text{Schw}}(\mu)$ is the standard Schwarzschild solution of mass $\mu$, and ${}_2 h_{ab}$ satisfies the vacuum Einstein equations linearized about the Schwarzschild solution. An explicit quadrupolar solution in Regge-Wheeler gauge is \cite{Detweiler:2001}
\beq
 {}_2 h_{ab} d\hat{x}^a d\hat{x}^b = -\hat{\mathcal{E}}_{ij} \hat{x}^i \hat{x}^j \left[  \left(1 - 2 \mu / \hat{r} \right)^2 d\hat{t}^2 + d\hat{r}^2 + (\hat{r}^2 - 2\mu^2) d\Omega^2 \right] + \frac{4}{3} \varepsilon_{kpq} \tensor{\hat{\mathcal{B}}}{^q _i} \hat{x}^p \hat{x}^i \left(1 - 2\mu / \hat{r} \right) d\hat{t} d\hat{x}^k .
\eeq
In a buffer region where $\mu / \hat{r} \rightarrow 0$ and $\hat{r} / M \rightarrow 0$ the `inner' and `outer' solutions mesh together. In Ref.~\cite{Detweiler:2001} Detweiler writes down an `overlap' solution of the form
\beq
g_{ab} \sim  \left( \bar{g}_{ab} + h^R_{ab} \right) + \left( g_{ab}^{\text{Schw}} + {}_2 h_{ab} \right) - \left(\eta_{ab} + {}_2 H_{ab} \right) + O(\mu^2 / M^2) .
\eeq
For $\hat{r} \ll M$ the first and third terms nearly cancel, leaving a tidally-perturbed Schwarzschild BH. For $\mu \ll \hat{r}$, the metric resembles $\bar{g}_{ab} + h^R_{ab} + (g_{ab}^{\text{Schw}} - \eta_{ab})$. Here, the final bracketed term is (a leading order approximation to) the Detweiler-Whiting singular field.

The key point in the argument sketched above is that, sufficiently close to the body of mass $\mu$, the physical metric resembles that of a tidally-perturbed black hole. The tidal perturbation is found by evaluating the electric-type and magnetic-type tidal tensors in the regular perturbed geometry $\bar{g}^R_{ab} + h^R_{ab}$ (i.e.~\emph{not} the full physical metric). Thus an observer in the vicinity of the body could, with a well-designed experiment, infer the tidal perturbation on the black hole that is induced by its motion through an external spacetime. For this reason, we should regard the shifts in the eigenvalues defined in previous sections as having a clear physical meaning. On the other hand, it should not be forgotten that local tidal effects in the vicinity of the mass $\mu$ will be dominated by the black hole itself (and not its tidal perturbation). Furthermore, if the body of mass $\mu$ is a compact body, which may carry an intrinsic quadrupole moment (e.g.~a neutron star), and change shape in response to external tides, then it may be much more difficult to separate external and local effects.

\subsubsection{Post-Newtonian expansion}
The argument sketched above could certainly be put on a more rigorous footing. One possibility would be to build on the work of Poisson on tidally-perturbed black holes in a light cone gauge \cite{Poisson:2003, Poisson:2005, Taylor:Poisson:2008}. In Ref.~\cite{Taylor:Poisson:2008}, Taylor \& Poisson have considered a tidally-perturbed black hole moving in an external geometry defined by a Post-Newtonian expansion. Implicit in Eq.~(1.10)--(1.16) of Ref.~\cite{Taylor:Poisson:2008} is an expansion of the tidal electric eigenvalues at 1PN relative order, and the magnetic eigenvalue at 0PN relative order. Johnson-McDaniel {\it et al.} \cite{JohnsonMcDaniel:2009} have gone further, by matching a PN metric to two tidally-perturbed Schwarzschild black holes. Implicit in Eq.~(B1a)--(B1b) of Ref.~\cite{JohnsonMcDaniel:2009} is the expansion of both electric and magnetic eigenvalues through 1PN relative order. In our notation,
\begin{eqnarray}
M^2 \, \lambda_1^E &=& -2 y^3 - 3 y^4 + \frac{\mu}{M} \left( 2 y^3 + 2 y^4 \right) + O(y^5) + O(\mu^2) ,
\label{eq:lambda1_PN}\\
M^2 \, \lambda_2^E &=&
\phantom{-2} y^3 + 3 y^4 + \frac{\mu}{M} \left( -y^3 - \frac{3}{2} y^{4} \right) + O(y^5) + O(\mu^2) ,
\\
M^2 \, \lambda_3^E &=&
\phantom{-2} y^3 + \; 0 \, \; + \frac{\mu}{M} \left( -y^3 - \frac{1}{2} y^4 \right) + O(y^5) + O(\mu^2) ,
\\
M^2 \, \lambda^B  &=& - 3 y^{7/2} - 6 y^{9/2} + \frac{\mu}{M} \left( 2 y^{7/2} + 3 y^{9/2} \right) + O(y^{11/2}) + O(\mu^2) ,
\label{eq:lambdaB_PN}
\end{eqnarray}
where $y = M / r_{\Omega}$.
Note that the $O(\mu^0)$ terms are Taylor-series expansions for the `test-particle' eigenvalues given in Sec.~\ref{subsec:testparticle}. The terms at $O(\mu^1)$ provide the leading terms in the PN expansions of $\Delta \lambda$. We will test these expansion against numerical results in Sec.~\ref{sec:results}.

\section{Method}\label{sec:method}

In this section we overview the calculation of the gauge-invariant quantities
$\Delta \psi$, $\Delta\lambda^{E/B}_i$ and $\Delta\chi$
 in the case of a particle moving on a circular orbit about a
Schwarzschild black hole. In the next section we will present our results. Our
calculation is made with two independent frequency-domain codes: i) a
Lorenz-gauge code implemented in C \cite{Akcay:Warburton:Barack:2013} and ii)
and a Regge-Wheeler-Zerilli (RWZ) gauge code implemented in {\sc{Mathematica}}.

Both codes decompose the metric perturbation into tensor spherical-harmonic and
frequency modes. For a generic setup the modes are indexed by the multipole
indices, $lm$, and the mode frequency $\omega$. In our case, as we are making
our calculation for circular orbits, $\omega=m\Omega$, only the $lm$
indices are required to label the modes. For each $lm$-mode appropriate boundary
conditions are imposed to solve for the retarded homogeneous metric perturbation.
The radiative, $m \neq 0$, modes of the metric perturbation are solved for
numerically. For the static, $m=0$, modes analytic solutions are known. The modes of the inhomogeneous metric perturbation are then constructed via the standard
variation of parameters method (as we have a delta-function source, this amounts
to imposing suitable jump conditions at the particle). Finally, for each
tensor-harmonic mode we project onto scalar harmonics, sum over $m$ and
regularize using the standard mode-sum approach \cite{Barack:Ori:1999}. The necessary regularization
parameters are given in Sec.~\ref{sec:mode-sum} below.

\subsection{Shift to asymptotically flat gauge\label{subsec:asymp-flat}}

In order to compare our results with PN theory it is necessary to work in an
asymptotically flat gauge. In both the Lorenz and Zerilli gauges the
$tt$-component of the metric perturbation does not vanish at spatial infinity
and so we make an $O(\mu)$ gauge transformation to correct for this
\cite{Sago:Barack:Detweiler:2008}. For both gauges this correction can be made
by adding $h^{NAF}_{ab} = \xi_{a ; b} + \xi_{b ; a}$ where
$\xi{^a} = [- \alpha(t + r_* - r), 0, 0, 0]$ and
$\alpha = \mu/\sqrt{r_0(r_0-3M)}$. Explicitly, this can be achieved by adding an
extra term to the invariants,
$\Delta \lambda \rightarrow \Delta \lambda + \delta^\xi \lambda$ where
\begin{subequations}
\begin{eqnarray}
  \delta^\xi \psi        &=& M \alpha / \sqrt{r_0 (r_0-3M)}, \\
  \delta^\xi \lambda^E_1 &=& -2 M \alpha(2r_0^2 - 8M r_0 + 9M^2)/(r_0^3(r_0-3M)^2), \\
  \delta^\xi \lambda^E_2 &=& 2M\alpha(r_0-2M)/(r_0^2(r_0-3M)^2), \\
  \delta^\xi \lambda^E_3 &=& 2M\alpha/r_0^3, \\
  \delta^\xi \lambda^B &=& -M^{3/2} \alpha (7 r_0^2 - 31 M r_0 + 36 M^2)/(r_0^3(r-3M)^2(r_0-2M)^{1/2}), \\
  \delta^\xi \chi &=& 0, \\
  \delta^\xi I &=& 0, \\
  \delta^\xi J &=& 0.
\end{eqnarray}
\end{subequations}

\subsection{Mode-sum regularization parameters}\label{sec:mode-sum}

In order to compute regularization parameters for the spin-precession
and tidal-tensor invariants, we require expressions for $\Delta\psi$,
$\Delta\lambda^E_i$ and $\Delta\lambda^B$ written in terms of the
components of $h_{ab}$ in Schwarzschild coordinates. There is a degree
of flexibility in the definition of $h_{ab}$ off the worldline; any
appropriately smooth extension off the worldline should suffice.
Here, we chose to work with an extension where the invariants take a
form which is convenient for computation, namely
\begin{align}
\label{eq:Delta-psi-Schwarzschild}
\Delta \psi = &
  \frac{1}{2 r_0 \Omega} \sqrt{\frac{r_0-3 M}{r_0}} \bigg[
      h_{tr,\phi}
    - h_{t\phi,r}
    + \Omega (h_{r\phi,\phi}-h_{\phi \phi,r} + f r_0 h_{rr})
    \bigg]
    \nonumber \\
    &
    +\frac{1}{2 M r_0 f} \sqrt{\frac{M}{r_0-3 M}} 
    \bigg[
      \Omega (
         M r_0^2 h_{tt}
        + r_0 f^2 h_{\phi \phi})
        + 2 M f h_{t\phi})
    \bigg],
\end{align}
\begin{align}
\label{eq:Delta-lambda1-Schwarzschild}
\Delta \lambda^E_1 = &
   \frac{\Omega^2 f (2 r_0 - 3 M)}{r_0 - 3 M} h_{rr}
   - \frac{\Omega^2(2 r_0^2-6 M r_0+3 M^2)}{f (r_0 - 3 M)^2} h_{tt}
   - \frac{6 M \Omega f h_{t\phi}}{r_0 (r_0 - 3 M)^2}
   - \frac{\Omega^2 (r_0^2-3 M r_0+3 M^2) h_{\phi\phi}}{r_0^2 (r_0 - 3 M)^2}
\nonumber \\
  &
 -\frac{r_0 - 2 M}{2 (r_0 - 3 M)} \Big[
     h_{tt,rr}
   + 2 \Omega h_{t\phi,rr}
   + \Omega^2 h_{\phi\phi,rr}
 \Big]
 -\frac{\Omega^2 h_{r\phi,\phi}
   + \Omega [h_{tr,\phi}+ h_{t\phi,r}]
   + h_{tt,r}}{r_0},
\end{align}
\begin{align}
\label{eq:Delta-lambda2-Schwarzschild}
\Delta \lambda^E_2 = &
  \frac{
    2 M [h_{t t} + 2\Omega h_{t \phi} + \Omega^2 h_{\phi \phi}]
    - [r_0 - 3 M] [
      h_{t t,\theta \theta}
      + 2\Omega h_{t \phi,\theta \theta}
      + \Omega^2 (h_{\phi \phi,\theta \theta} + 2 h_{\theta \theta})
    ]\sin^2 \theta }{2r_0 (r_0 - 3 M)^2} ,
\end{align}
\begin{align}
\Delta \lambda^E_3 =
  \frac{\Omega^2}{f} h_{t t}
  -\Omega^2 f h_{r r}
  -\frac{\Omega^2}{r_0^2} h_{\phi \phi}
  +\frac{\Omega (h_{t \phi, r} - h_{t r, \phi})
    + \Omega^2(h_{\phi \phi, r} - h_{r \phi, \phi})}{r_0}
  -\frac{h_{t t, \phi \phi}
    + 2\Omega h_{t \phi, \phi \phi}
    + \Omega^2 h_{\phi \phi, \phi \phi}}{2 r_0^2 f},
\nonumber \\
\end{align}
\begin{align}
\label{eq:Delta-lambda3-Schwarzschild}
\Delta \lambda^B &=
 \frac{3\Omega^3 f^{1/2}\sin^2\theta}{(r_0-3 M)} h_{\theta\theta}
+\frac{\Omega^3 f^{1/2}(r_0-9 M)}{2(r_0-3 M)^2}h_{\phi\phi}
-\frac{\Omega^2 (r_0-M) h_{t\phi}}{\sqrt{f} (r_0-3 M)^2}
-\frac{\Omega M (5 r_0-9 M) h_{tt}}{2 \sqrt{f} r_0 (r_0-3 M)^2}
-\frac{\Omega f^{3/2}}{r_0 } h_{rr}
\nonumber \\
&
+\frac{\sqrt{f}}{2 r_0^2} \Big[
  \Omega [\sin^2 \theta (h_{\theta\theta,r}
  -2 h_{r\theta,\theta})
  -h_{r\phi,\phi}]-h_{tr,\phi}
\Big]
+\frac{1}{2 \sqrt{f} r_0^3}\Big[(r_0-4 M) h_{t\phi,r} +\Omega (r_0-3 M) h_{\phi\phi,r}\Big]
-\frac{\Omega h_{tt,r}}{2 \sqrt{f}}
\nonumber \\
&
+\frac{\Omega \sin^2 \theta}{2 \sqrt{f} r_0^2 (r_0-3 M)}\Big[
  f h_{\phi\phi,\theta\theta}+r_0^2 h_{tt,\theta\theta}\Big]
-\frac{\sin^2\theta}{2 \sqrt{f} r_0^3} \Big[
  \Omega  h_{\theta\phi,\phi\theta}+h_{t\theta,\phi\theta}\Big]
+\frac{(r_0-M)\sin^2\theta}{2 \sqrt{f} r_0^3 (r_0-3 M)} h_{t\phi,\theta\theta}.
\end{align}
Using this definition, the regularization parameters may then be derived
using the methods of Ref.~\cite{Heffernan:Ottewill:Wardell:2012} to decompose
into scalar spherical harmonics. Doing so, we obtain a mode-sum formula for each
of the invariants of the form
\begin{equation}\label{eq:mode_sum}
  \Delta \lambda^R = \eta \sum_{\ell=0}^{\infty} \Big[\Delta \lambda_{\ell}^{\rm ret}
    - (2\ell+1)^2 \Delta \lambda_{[-2]} - (2\ell+1) \Delta \lambda_{[-1]}
    - \Delta \lambda_{[0]}\Big]
\end{equation}
where the coefficients for each of the invariants are given by\footnote{These
may be downloaded in electronic form as a \textsc{Mathematica} notebook 
\cite{barrywardellnet}.}
\begin{equation}
  \Delta \psi_{[-2]} = 0,
\quad
  \Delta \psi_{[-1]} = \mp \frac{r_0-3M}{2 r_0 (r_0-2 M)},
\quad
  \Delta \psi_{[0]} = \frac{(r_0-3M) [ (9 M - 4 r_0) \mathcal{E} + 2 \, (2 r_0 - 5 M) \mathcal{K} ]}{M \pi \sqrt{r_0^3 (r_0-2M)}},
\end{equation}
\begin{equation*}
  \Delta \lambda_{1\,[-2]} = -\frac{M \mathcal{E}}{2 \pi  r_0^3} \sqrt{\frac{r_0-3 M}{r_0-2 M}},
\quad
  \Delta \lambda_{1\,[-1]} = \mp \frac{M^2 \sqrt{r_0-3 M}}{r_0^{7/2} (r_0-2 M)},
\end{equation*}
\begin{equation}
  \Delta \lambda_{1\,[0]} = \frac{M [ (23 r_0^2-91 M r_0+82 M^2) \mathcal{E} - 3 (7 r_0^2-38 M r_0+35 M^2) \mathcal{K} ]}{4 \pi  r_0^4 \sqrt{(r_0-2M)(r_0-3M)}},
\end{equation}
\begin{equation*}
  \Delta \lambda_{2\,[-2]} = \frac{1}{2 \pi  r_0^3} \sqrt{\frac{r_0-3 M}{r_0-2 M}} \Big[ \mathcal{E} (r_0-2M)-\mathcal{K} (r_0-3M) \Big],
\quad
  \Delta \lambda_{2\,[-1]} = 0,
\end{equation*}
\begin{align}
  \Delta \lambda_{2\,[0]} =&
  \frac{
    \mathcal{E} (r_0-2 M) (16 r_0^2 + 45 M r_0 - 199 M^2 )
    -2\mathcal{K} (8 r_0^3 - M r_0^2 -144 M^2 r_0 + 249 M^3)
  }{4 \pi  r_0^4 \sqrt{(r_0-2M)(r_0-3M)}} ,
\end{align}
\begin{equation*}
  \Delta \lambda_{3\,[-2]} = \frac{(r_0-3M)^{3/2}}{2 \pi  r_0^3 \sqrt{r_0-2M}}(\mathcal{K} - \mathcal{E}),
\quad
  \Delta \lambda_{3\,[-1]} = \pm\frac{M^2 \sqrt{r_0-3M}}{r_0^{7/2} (r_0-2 M)},
\end{equation*}
\begin{equation}
  \Delta \lambda_{3\,[0]} = \sqrt{\frac{r_0-3 M}{r_0-2 M}} \frac{1}{4 \pi  r_0^4}\Big[4 \mathcal{E} (40 M^2-29 M r_0+4 r_0^2)-\mathcal{K} (201 M^2-123 M r_0+16 r_0^2) \Big],
\end{equation}
\begin{equation*}
  \Delta \lambda^B_{[-2]} = 0, \quad
  \Delta \lambda^B_{[-1]} = \pm \frac{M}{2 r_0^2} \sqrt{\frac{M}{r_0^3}}\left(\frac{r_0-3 M}{r_0-2 M}\right)^{3/2},
\end{equation*}
\begin{align}
  \Delta \lambda^B_{[0]} =& \frac{1}{\pi r_0^4 (r_0-2 M)(r_0-3 M)^{1/2}} \Big[
  2 (75 M^4-119 M^3 r_0+71 M^2 r_0^2-19 M r_0^3+2 r_0^4) \mathcal{K}
\nonumber \\
  & \qquad - (129 M^4-206 M^3 r_0+127 M^2 r_0^2-36 M r_0^3+4 r_0^4)\mathcal{E} \Big],
\end{align}
\begin{equation}
  \Delta \chi_{[-2]} = 0, \quad
  \Delta \chi_{[-1]} = 0, \quad
  \Delta \chi_{[0]} = 0. \quad
\end{equation}
Here,
\begin{equation}
\mathcal{K} \equiv \int_0^{\pi/2} \bigg(1 - \frac{M \sin^2 \theta}{r_0-2M}\bigg)^{-1/2} d\theta, \quad
\mathcal{E} \equiv \int_0^{\pi/2} \bigg(1 - \frac{M \sin^2 \theta}{r_0-2M}\bigg)^{1/2} d\theta
\end{equation}
are complete elliptic integrals of the first and second kinds, respectively.
It is also possible to add higher-order terms to Eq.~\eqref{eq:mode_sum} to increase the rate of convergence of the mode-sum with $l$ \cite{Heffernan:Ottewill:Wardell:2012}. With the regularization parameters given above the contribution to the mode-sum for $\Delta\lambda^{E/B}_i$ goes as $1/l^{2}$ for high $l$. For $\Delta\chi$ all the regularization parameters are zero and the mode-sum converges exponentially.

Note that $\Delta \lambda_{1,\,[-2]} + \Delta \lambda_{2,\,[-2]} +\Delta \lambda_{3,\,[-2]}  = 0$
and $\Delta \lambda_{1,\,[-1]} + \Delta \lambda_{2,\,[-1]} +\Delta \lambda_{3,\,[-1]}  = 0$,
as expected. While this does not hold for our expression for
$\Delta \lambda_{1,\,[0]} + \Delta \lambda_{2,\,[0]} +\Delta \lambda_{3,\,[0]}$,
this is not a reflection of an error in either our expressions for the
regularization parameters, or the tracelessness of the perturbed eigenvalues,
Eq.~\eqref{eq:traceless-perturbed-eigenvalues}. Instead, it is merely a
reflection of the particular choice of off-worldline extension of $h_{ab}$ that
we made in computing the expressions for
Eqs.~\eqref{eq:Delta-lambda1-Schwarzschild},
\eqref{eq:Delta-lambda3-Schwarzschild} and
\eqref{eq:Delta-lambda3-Schwarzschild}. It is therefore important to use the
same expressions to construct the $\Delta \lambda^{\rm ret}_{\ell}$ from the
retarded metric perturbation. Importantly, the regularized sum,
$\Delta \lambda^R$ is not modified by this choice of off-worldline extension
and we find that
$\Delta \lambda_1^R + \Delta \lambda_2^R + \Delta \lambda_3^R = 0$, as expected.

\section{Results\label{sec:results}}

\subsection{Data and figures}
Table \ref{table:results} presents accurate numerical results for the four independent gauge-invariant tidal degrees of freedom at $O(\mu)$ associated with quasi-circular orbits of a Schwarzschild black hole, for orbital radii in the range $4M \le r_0  \le 5000 M$.

\begin{figure}
 \includegraphics[width=12cm]{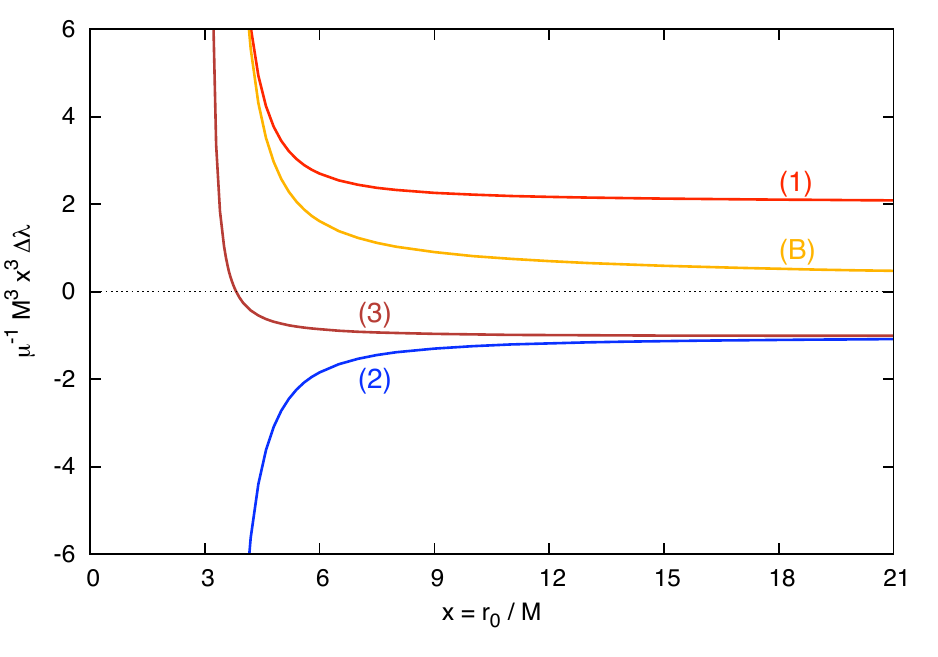}
 \caption{
  Perturbation in eigenvalues of tidal tensors, $\{ \Delta \lambda_1^E, \Delta \lambda_2^E, \Delta \lambda_3^E, \Delta \lambda^B  \}$ [defined by Eqs.~(\ref{dlam}) and (\ref{dellam1})--(\ref{dellamB})] at $O(\mu)$, for a quasi-circular geodesic on Schwarzschild at frequency-radius $r_\Omega$. Note that the eigenvalues are here scaled by $(r_0 / M)^3$, and that $\Delta \lambda_3^E$ changes sign around $r_0\approx 3.802 M$.
 } \label{fig:eigs_strongfield}
\end{figure}

As shown in Fig.~\ref{fig:eigs_strongfield}, we find that $\Delta \lambda^E_1$ and $\Delta \lambda^B$ are positive and monotonically decrease with increasing $r_0$. Similarly, $\Delta \lambda^E_2$ is negative and monotonically increases with increasing $r_0$. The third electric-type eigenvalue, $\Delta \lambda^E_3$, exhibits more structure with a zero crossing near the light-ring; we find $\Delta \lambda^E_3$ is negative for $r_0\gtrapprox 3.802M$ and positive otherwise. 

\begin{table}
\begin{center}
\begin{tabular}{c | c  c  c | c }
\hline\hline
$r_\Omega/M$ & $\tilde{\Delta} \lambda^E_1$ & $\tilde{\Delta} \lambda^E_2$ & $\tilde{\Delta} \lambda^B$ & $\Delta\chi $ \\
\hline
$4$	& $1.246430830\times 10^{-1}$	& $-1.2073036467\times 10^{-1}$	& $1.1950561710\times 10^{-1}$	& $6.20599279790\times 10^{-2}$	\\
$5$	& $2.76048162228\times 10^{-2}$	& $-2.17623317583\times 10^{-2}$	& $2.06379254102\times 10^{-2}$	& $3.07987615215\times 10^{-2}$	\\
$6$	& $1.25009071026\times 10^{-2}$	& $-8.53346795295\times 10^{-3}$	& $7.46326072544\times 10^{-3}$	& $1.80359457690\times 10^{-2}$	\\
$7$	& $7.13119569832\times 10^{-3}$	& $-4.46794557538\times 10^{-3}$	& $3.59482492474\times 10^{-3}$	& $1.16897679241\times 10^{-2}$	\\
$8$	& $4.54662923752\times 10^{-3}$	& $-2.70175256799\times 10^{-3}$	& $2.01488674141\times 10^{-3}$	& $8.10949353524\times 10^{-3}$	\\
$9$	& $3.10253317396\times 10^{-3}$	& $-1.78069581806\times 10^{-3}$	& $1.24157717625\times 10^{-3}$	& $5.90798502939\times 10^{-3}$	\\
$10$	& $2.21987210893\times 10^{-3}$	& $-1.24378422421\times 10^{-3}$	& $8.16987907232\times 10^{-4}$	& $4.46668994779\times 10^{-3}$	\\
$12$	& $1.25561205099\times 10^{-3}$	& $-6.82302030854\times 10^{-4}$	& $4.04957204672\times 10^{-4}$	& $2.77073054838\times 10^{-3}$	\\
$14$	& $7.79965771010\times 10^{-4}$	& $-4.16146336672\times 10^{-4}$	& $2.27128540894\times 10^{-4}$	& $1.85939663738\times 10^{-3}$	\\
$16$	& $5.17690252913\times 10^{-4}$	& $-2.72911197989\times 10^{-4}$	& $1.38691988691\times 10^{-4}$	& $1.31984784139\times 10^{-3}$	\\
$18$	& $3.61123864339\times 10^{-4}$	& $-1.88776522748\times 10^{-4}$	& $9.01546982691\times 10^{-5}$	& $9.77171317875\times 10^{-4}$	\\
$20$	& $2.61878410440\times 10^{-4}$	& $-1.36049623949\times 10^{-4}$	& $6.14931781943\times 10^{-5}$	& $7.47601173528\times 10^{-4}$	\\
$30$	& $7.64205652554\times 10^{-5}$	& $-3.90707419872\times 10^{-5}$	& $1.43357179908\times 10^{-5}$	& $2.68628079813\times 10^{-4}$	\\
$40$	& $3.19976522699\times 10^{-5}$	& $-1.62499773274\times 10^{-5}$	& $5.15281649939\times 10^{-6}$	& $1.30511975094\times 10^{-4}$	\\
$50$	& $1.63081337530\times 10^{-5}$	& $-8.25202517458\times 10^{-6}$	& $2.33815072508\times 10^{-6}$	& $7.46610965586\times 10^{-5}$	\\
$60$	& $9.40858534234\times 10^{-6}$	& $-4.75000366923\times 10^{-6}$	& $1.22795590189\times 10^{-6}$	& $4.73327668903\times 10^{-5}$	\\
$70$	& $5.91181734892\times 10^{-6}$	& $-2.98000495794\times 10^{-6}$	& $7.13003338522\times 10^{-7}$	& $3.22053620959\times 10^{-5}$	\\
$80$	& $3.95382643810\times 10^{-6}$	& $-1.99078798763\times 10^{-6}$	& $4.45463263366\times 10^{-7}$	& $2.30739345760\times 10^{-5}$	\\
$90$	& $2.77326058007\times 10^{-6}$	& $-1.39517219566\times 10^{-6}$	& $2.94289927146\times 10^{-7}$	& $1.71958426589\times 10^{-5}$	\\
$100$	& $2.01957688484\times 10^{-6}$	& $-1.01533032618\times 10^{-6}$	& $2.03156640710\times 10^{-7}$	& $1.32193326253\times 10^{-5}$	\\
$500$	& $1.60318513516\times 10^{-8}$	& $-8.02409469186\times 10^{-9}$	& $7.17709743776\times 10^{-10}$	& $2.37840715576\times 10^{-7}$	\\
$1000$	& $2.00199530253\times 10^{-9}$	& $-1.00150291700\times 10^{-9}$	& $6.33408908079\times 10^{-11}$	& $4.20979092138\times 10^{-8}$	\\
$5000$	& $1.60031984834\times 10^{-11}$	& $-8.00240092269\times 10^{-12}$	& $2.26342119063\times 10^{-13}$	& $7.53980143152\times 10^{-10}$	\\
\hline \hline
\end{tabular}
\end{center}
\caption{Numerical results for tidal invariants $\Delta\lambda^E_1$, $\Delta\lambda^E_2$, $\Delta\lambda^B$ and $\Delta\chi$. The third electric-type eigenvalue, $\Delta\lambda^E_3$, can be constructed from the first two using the traceless condition $\Delta\lambda^E_1+\Delta\lambda^E_2+\Delta\lambda^E_3=0$. We believe that all digits presented are accurate. Here, $\tilde{\Delta} \lambda$ indicates the dimensionless version, $\mu^{-1} M^3 \Delta \lambda$. [{\it Note added (Dec 2014)}: The data in this table has been corrected. As highlighted in Fig.~1 of Ref.~\cite{Bini:Damour:2014c}, the original data set for $\tilde{\Delta} \lambda_i^{E/B}$ was afflicted by a small but unanticipated error with a relative magnitude below $10^{-5}$. The error was traced to a bug in the implementation of numerical fits to high-order regularization parameters.]
}\label{table:results}
\end{table}

\begin{figure}
 \includegraphics[width=12cm]{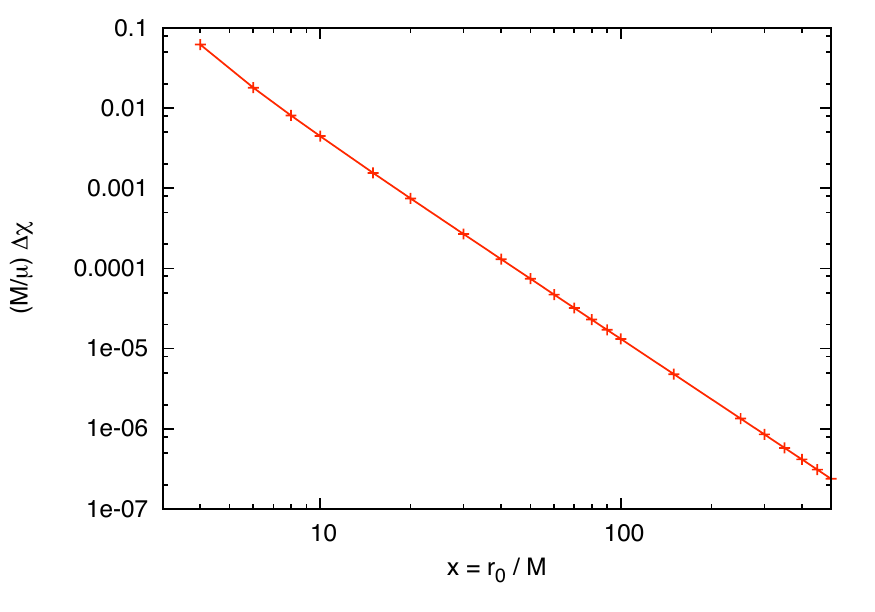}
 \caption{
  Angle $\chi$ defined by electric and magnetic eigenvectors, Eqs.~(\ref{eq:chidef}) and (\ref{eq:dchi}), for quasi-circular orbit on Schwarzschild at $O(\mu)$. The plot shows $\Delta \chi$ as a function of the orbital radius. Note that $\chi$ is dissipative in character. In the far-field, $(M / \mu) \Delta \chi \approx \frac{4}{3} y^{-5/2} - \frac{13}{5} y^{-7/2}$ where $y = M / r_0$. } \label{fig:chi}
\end{figure}

\begin{figure}
 \includegraphics[width=12cm]{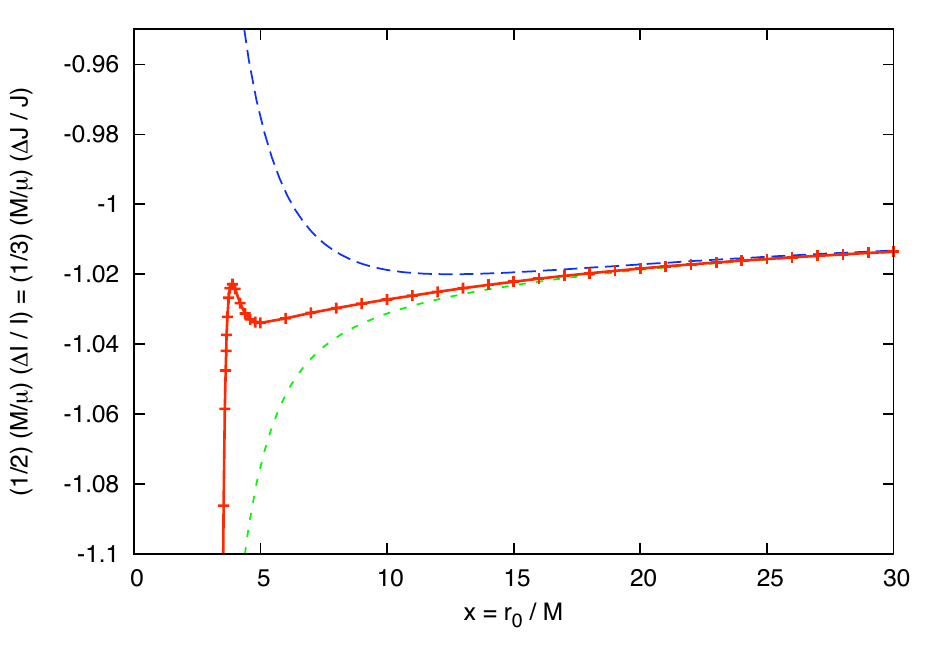}
 \caption{
 Perturbation in curvature scalars on the quasi-circular orbit on Schwarzschild at $O(\mu)$. The plot shows numerical data [red, solid] for the relative shifts in the Kretschmann and third-order scalars,  $\frac{1}{2} \frac{\Delta I}{I}$ and $\frac{1}{3} \frac{\Delta J}{J}$. Note that at $O(\mu)$ these shifts are equivalent due to algebraic speciality of the background spacetime, which implies that the speciality index is $\mathcal{S} = 1 + O(\mu^2)$. See Sec.~\ref{subsec:curvature-invariants}) for details. The dotted lines show successive post-Newtonian series, $p(y) = -1 - \frac{1}{2} y + \frac{25}{8} y^2$ [blue dashed] and $p(y)  - \frac{25}{2} y^3$ [green dotted], where $y = M / r_0$. Coefficients at orders $y^2$ and above have been inferred from a numerical fitting. 
 } \label{fig:curvature}
\end{figure}

\begin{figure}
 \includegraphics[width=12cm]{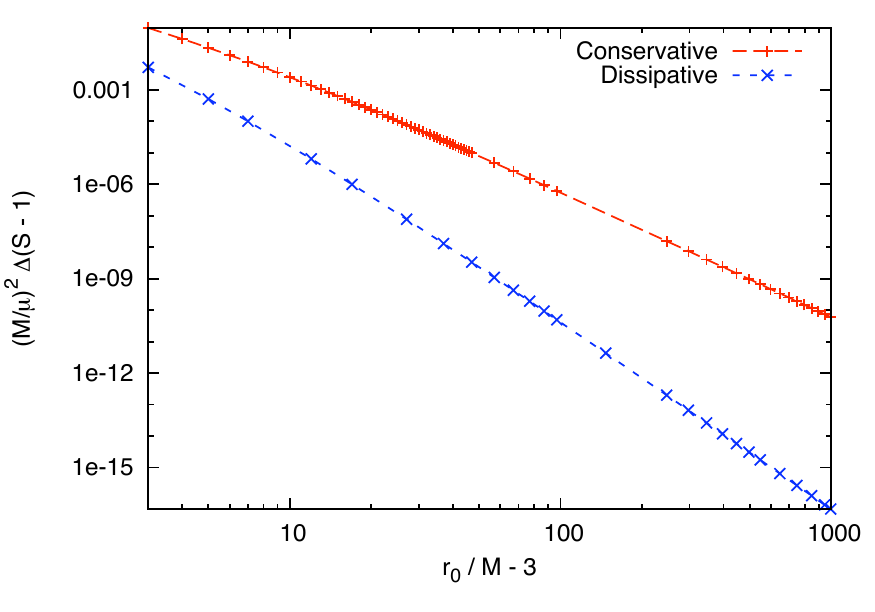}
 \caption{
Perturbation in the speciality index $\mathcal{S} = 27 J^2 / I^3$, evaluated on the quasi-circular orbit on Schwarzschild at $O(\mu^2)$. The plot shows numerical data for the perturbation in the speciality index, $ \mathcal{S} - 1$, at $O(\mu^2)$, in the regular perturbed spacetime, calculated via Eq.~(\ref{eq:DeltaS}). This represents an invariant measure of the change in Petrov type, from background Type D (algebraically special) to perturbed Type I. In the far-field the conservative and dissipative contributions scale as $\sim -\frac{243}{4} \, y^4$ and $\sim 48 \, y^6$, respectively, where $y = M / r_0$.
 } \label{fig:speciality}
\end{figure}

In Figs.~\ref{fig:eigs_strongfield}--\ref{fig:speciality} we plot the various tidal invariants as a function of the circular orbit radius. The behaviour in the weak-field and near the light-ring is explored in more detail in the following sections. 

Figure \ref{fig:chi} shows that the dissipative quantity $\Delta \chi$, which is defined in terms of an angle between electric and magnetic eigenvectors, is a monotonically increasing function with apparently no additional structure. Figure \ref{fig:curvature} shows the relative shift in the second- and third-order curvature scalars, which at $O(\mu)$ are not linearly independent (see Eq.~(\ref{eq:dI_over_I})). Intriguingly, there appears a local minimum and local maximum in the very strong field regime, somewhat before the light-ring, which may perhaps affect the convergence of PN series. The local maximum is at radius somewhat close to the zero-crossing of $\Delta \lambda^E_3$.

Figure \ref{fig:speciality} shows the deviation of the speciality index $\mathcal{S}$ from unity at $O(\mu^2)$. Note that $\Delta \mathcal{S}$ is constructed from quadratic combinations of $O(\mu)$ quantities, via Eq.~(\ref{eq:DeltaS}). It has a `conservative' part given in terms of $\Delta \lambda$ and a `dissipative' part given in terms of $\Delta \chi$, with quite different leading-order scalings in $M/r_0$. The plot shows that, unlike the background spacetime, the perturbed spacetime is \emph{not} Petrov Type D. The deviation from speciality increases monotonically as the orbital radius decreases.

In addition we have calculated Detweiler's redshift invariant and the spin invariant, the results of which we give in Appendix \ref{apdx}.

\subsection{Numerical accuracy}
For the computation of the gauge-invariant quantities the results of our Lorenz-gauge code are accurate to $7$--$8$ significant figures in the range $r_0=4$---$100M$. By contrast, the RWZ code is accurate to about $12$--$13$ significant figures in the range $r_0=4$---$5000M$. The results of both codes agree to
within the error bars of the Lorenz-gauge code for $r_0\le100M$. The more accurate results from our RWZ code are the ones presented in Table \ref{table:results}. 

It is interesting to note that both our Lorenz-gauge and RWZ codes produce higher accuracy results when computing $\Delta U$ or $\Delta \psi$. As an example, by comparison with known high-order PN results \cite{Bini:Damour:2014}, our RWZ code computes $\Delta U$ at $r_0 = 5000M$ to 18 significant figures ({\sc{Mathematica}} allows us to go beyond machine precision in our calculations with ease --- see Appendix \ref{apdx}). Similarly, our RWZ codes computes $\delta\psi$ to 15 significant figures at $r_0=5000M$. The reason for this range in accuracy when computing the different gauge-invariants is two-fold. Firstly, for asymptotically high $l$ the individual $l$-modes of the retarded and singular field for $\Delta U$, $\Delta \psi$ and $\Delta \lambda^{E/B}_i$ go as $l^0,l^1,$ and $l^2$ respectively. Secondly, the leading-order PN contributions are $r_0^{-1}$, $r_0^{-2}$ and $r_0^{-3}$ respectively. Hence, for example, when calculating $\Delta U$ we must subtract (for large $l$) two small quantities to find a large one. By contrast, when calculating the tidal-tensor eigenvalues we must subtract (for large $l$) two large quantities to get a relatively small one. This requirement to calculate a small quantity buried in the difference between two large quantities is the reason for the difference in accuracy when calculating the different gauge invariants.

\subsection{Fitting for unknown coefficients in the PN series\label{subsec:PNfitting}}

The high accuracy of our numerical data out as far as $r_0=5000M$ allows us to fit for the currently unknown coefficient in the PN expansion. A similar program was undertaken for Detweiler's redshift invariant, $\Delta U$, by Blanchet \etal~\cite{Blanchet:2010} and Shah \etal~\cite{Shah:Friedman:Whiting:2013}, with their results later confirmed by the analytic calculations of Bini and Damour \cite{Bini:Damour:2013,Bini:Damour:2014a}.

In fitting for the coefficients of the PN series we use 25 data points with $r_0\ge100$ and assume that the PN series takes the form:
\begin{align}
	 \Delta \lambda^E_i (y \ll 1) = \frac{\mu}{M^3} \sum_{n=3}^\infty(a_n^i + b_n^i \ln(y))y^n				\label{eq:lambda_E_fit}	\\
	 \Delta \lambda^B_i (y \ll 1) = \frac{\mu}{M^3} \sum_{n=3}^\infty(a_n^B + b_n^B \ln(y))y^{n+1/2}			\label{eq:lambda_B_fit}
\end{align}
where $y=M/r_0$ and $n \in \mathbb{Z}$ for $n\le 5$. For $n>5$ we allow integer and half-integers values in the series. This form of the PN series is inspired by the known forms for $\Delta U$ \cite{Bini:Damour:2014a} and $\Delta \psi$ \cite{Bini:Damour:2014}. We fit and analyse our data using the {\sc{LinearModelFit}} package of {\sc{Mathematica}}. We find agreement to greater than 10 significant figures with the leading and sub-leading terms in the PN series presented in Eqs.~\eqref{eq:lambda1_PN}-\eqref{eq:lambdaB_PN}. We proceed by subtracting these terms from our data and fitting for the next few unknown coefficients. Our results are presented in Table \ref{table:fitted_PN_coeffs} and suggest the following terms are exactly:
\begin{align}
	a_5^1 = -\frac{19}{4},	\qquad a_5^2 = -\frac{23}{8},\qquad a_5^3 = \frac{61}{8} \qquad a_5^B = \frac{59}{4},	\label{eq:PN_fit_coeffs}	\\
	b_5^1 = 0,\qquad b_5^2 = 0,\qquad b_5^3 = 0,\qquad b_5^B = 0.
\end{align}
Unlike $\Delta U$ and $\delta\psi$ we find no evidence for a $\log$-term at relative 2PN order. We have also independently fitted for the coefficients in the PN expansion of the invariant defined in Eq.~\eqref{eq:dI_over_I}. We find
\begin{align}
	\left(\frac{1}{2}\frac{\Delta I}{I}=\frac{1}{3}\frac{\Delta J}{J}\right) (y \ll 1) = -1 - \frac{1}{2} y + \frac{25}{8}y^2 - 12.504(5)y^3.
\end{align}
The leading and sub-leading terms come from the known expansions of the tidal-tensor eigenvalues. Our fit suggests that the coefficient of $y^3$ is exactly -25/2, which is consistent with the fitted coefficients for the tidal-tensor eigenvalues in Eqs.~\eqref{eq:PN_fit_coeffs}.

We can also fit for the coefficients in the PN expansion of $\Delta\chi$. We are not (at present) aware of any analytic calculation of the leading-order terms and so we must fit for these as well. For small $y$ we find
\begin{align}
	\Delta\chi(y\ll1) = 1.3333335(6)y^{5/2} - 2.6002(5)y^{7/2} + 17.33(3) y^{4},
\end{align}
which suggests that the coefficients of the leading and sub-leading terms are exactly $4/3$ and $-13/5$, respectively.

\begin{table}
\begin{center}
\begin{tabular}{c | c c c c c c c c}
\hline\hline
	  & $a_n^1$ 		& $a_n^2$ 		& $a_n^3$ 		& $a_n^B$ 			& $b_n^1$ & $b_n^2$ & $b_n^3$ &$b_n^B$ \\
\hline
$n=3$ & $2$ 			& $-1$ 			& $-1$			& $2$				& 0								&	0								&	0								& 	0									\\
$n=4$ & $2$ 			& $-3/2$ 		& $-1/2$		& $3$	& 0								&	0								&	0								& 	$1^{+7}_{-5}\times10^{-6}$			\\
$n=5$ & $-4.7499(7)$	& $-2.8750(4)$	& $7.6249(5)$	& $14.7499(6)$		& $6^{+61}_{-87}\times10^{-6}$	&	$-1^{+25}_{-26}\times10^{-5}$	&	$-5^{+48}_{-49}\times10^{-6}$	& $-3^{+555}_{-562}\times10^{-7}$		\\
\hline \hline
\end{tabular}
\end{center}
\caption{Fitted coefficients of the PN series for the tidal-tensor eigenvalues -- see Eqs.~\eqref{eq:lambda_E_fit} and \eqref{eq:lambda_B_fit} for the form of the series being fit to. Numbers in brackets show the estimated error in the final digit, i.e., $-4.7499(7)=4.7499\pm0.0007$. 
}\label{table:fitted_PN_coeffs}
\end{table}

\subsection{Informing EOB theory\label{subsec:EOB}}
Using the above results, we may also infer PN expansions for quantities relevant to EOB theory. For example, Ref.~\cite{Bini:2012gu} highlights the role of (among other things) the `electric-quadrupole' invariant $\EE^2$ in the tidal action of EOB theory, defined by
\beq
\EE^2 \equiv \EE_{ab} \EE^{ab} = (\lambda_1^E)^2 + (\lambda_2^E)^2 + (\lambda_3^E)^2 .
\eeq
From our results, we can compute $\EE^2$ through $O(\mu)$, using $
\Delta \EE^2 \equiv 2 \left( \bar{\lambda}_1^E \Delta \lambda_1^E + \bar{\lambda}_2^E \Delta \lambda_2^E  + \bar{\lambda}_3^E \Delta \lambda_3^E \right)$. From our data, we infer the following PN expansion,
\beq
\Delta \EE^2  = -12 y^6 - 30 y^7 - \frac{93}{2} y^8 + \ldots
\eeq
Here, the first two terms are consistent with the expansion given in Eq.~(4.14) of Ref.~\cite{Bini:2012gu} (bearing in mind that $r_{12}$ and $r_{\Omega}$ should be related at $O(\mu)$ using Eq.~(4.12) in Ref.~\cite{Bini:2012gu} for the orbital frequency). The third term represents a prediction of the coefficient at next order. In addition, our numerical results can provide information on the \emph{global} behaviour of $\EE^2$ through $O(\mu)$, all the way up to the light ring.

\subsection{Behaviour near the light-ring}

In order to produce global fits for gauge-invariant quantities that can be used, for instance, to constrain free functions in EOB theory is it necessary to understand the behaviour of the relevant quantities as the orbital radius approaches to the light-ring. Akcay \etal~\cite{Akcay:Barack:Damour:Sago:2012} carried out the first such analysis with $h^{R,F}_{uu}\equiv h^{R,F}_{ab}u^a u^b$, a quantity related to the redshift invariant $\Delta U$. Here the superscripts $R$ and $F$ denote `regular' and `flat' respectively, with the latter implying the quantity is computed in an asymptotically flat gauge. In Ref.~\cite{Akcay:Barack:Damour:Sago:2012} $h^{R,F}_{uu}$ was found to diverge as $0.280(1)z^{-3/2}$, where $z=1-3M/r_0$. Bini and Damour \cite{Bini:Damour:2014} have also considered the divergence of the spin-precession invariant, $\Delta\psi$, at the light-ring and used this knowledge, along with their analytically derived high-order PN expansion, to further inform EOB theory \cite{Bini:Damour:2014}. They argue, based the known rate of divergence of $h^{R,F}_{uu}$, that $\Delta\psi$ will diverge like $0.1041(1)z^{-1}$.

\begin{figure}
	\includegraphics[width=10cm]{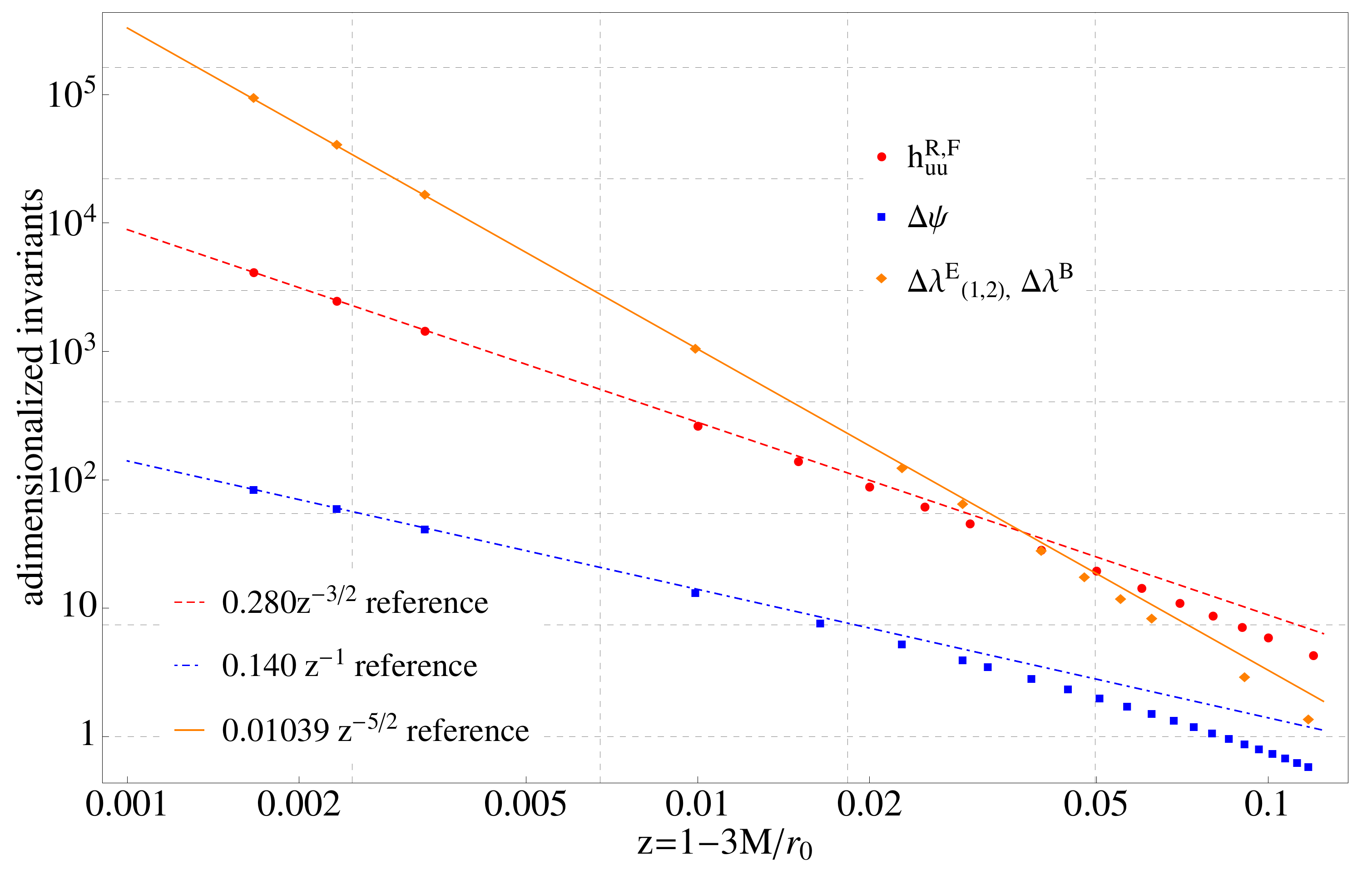}
	\caption{The divergence of the various gauge-invariants as the light ring at $r_0=3M$ $(z=0)$ is approached. The redshift invariant is related to $h^{R,F}_{uu}$ which is known to diverge as $z^{-3/2}$ \cite{Akcay:Barack:Damour:Sago:2012}. Bini and Damour \cite{Bini:Damour:2014} argued from knowledge of the behaviour of $h^{R,F}_{uu}$ at the light-ring that $\Delta\psi$ would diverge as $-0.1401(1)z^{-1}$. Here we confirm their prediction. Lastly, we find that $|\Delta\lambda^E_{1,2}|$ and $\Delta\lambda^B$ diverge as $\sim0.01039z^{-5/2}$. Our data is not sufficiently accurate to determine the sub-dominant rate of divergence of $\Delta\lambda^E_3$.}\label{fig:light-ring}
\end{figure}

In this section we present results for the divergence of a number of gauge invariants as the light-ring is approached. Our main results are encapsulated in Fig.~\ref{fig:light-ring}. For $h_{uu}^{R,F}$ we verified the leading-order divergence found by Akcay \etal~as $z\rightarrow0$. We have also extend our data to orbits closer to the light-ring than they were able to achieve which is particularly important for ascertaining the rate of divergence of the other gauge-invariant quantities. For $\Delta\psi$ we have confirmed the prediction of Bini and Damour for the leading-order divergence.

For the tidal-tensor eigenvalues $\Delta\lambda^E_{(1,2)}$ and $\Delta\lambda^B$ we find the three quantities diverge like $0.01039(5)z^{-5/2}$, $-0.01039(2)z^{-5/2}$ and $0.01039(1)z^{-5/2}$, respectively, as the light-ring is approached. Our data is not sufficiently accurate to determine the rate of divergence of $\Delta\lambda^E_3$. To understand why recall that the sum of the three electric-type eigenvalues is zero. Our results suggest that at leading order, the first two diverge at the same rate, but with opposite signs, and the third is (minus) the sum of these two. The value of $\Delta\lambda^E_3$ thus becomes ever more difficult to resolve as the light-ring is approached. With our current dataset the best we can say is that rate of divergence of $\Delta\lambda^E_3$ is sub-dominant to the other two electric-type eigenvalues.



\section{Discussion and conclusion\label{sec:discussion}}

In the preceding sections we have attempted to extract all physical content associated with a quasi-circular geodesic in a regular perturbed spacetime equipped with an equatorial symmetry, when one restricts attention to second (and lower) derivatives of the metric. For the case of equatorial circular orbits, we found, in addition to the `redshift' and `spin-precession' quantities, four independent `tidal' degrees of freedom. Namely, three independent eigenvalues (i.e.~the electric $\{ \lambda_1^E$, $\lambda_2^E$, $\lambda^E_3 = -\lambda^E_1 - \lambda^E_2 \}$ and magnetic $\{ \lambda^B , - \lambda^B, 0 \}$ sets) and one angle $\chi$, formed from a scalar product of electric and magnetic eigenvectors. The former are conservative in character, whereas the latter is dissipative. We have computed these quantities at $O(\mu)$ in Lorenz and RWZ gauges, verifying their gauge invariance. In Table \ref{table:results} we gave a sample of highly-accurate numerical results. From the four independent invariants, we are able to compute (on the worldline) additional quantities including the curvature scalars at $O(\mu)$, and the speciality index $\mathcal{S}$ at $O(\mu^2)$.

This work opens up several avenues for investigation. First, we anticipate that high-order PN expansions of the four invariant degrees of freedom described here can be obtained by following the approach pioneered by Bini \& Damour \cite{Bini:Damour:2013, Bini:Damour:2014}, which employs the formalism of Mano, Suzuki and Takasugi \cite{Mano:1996vt}. Second, after examining the behaviour at the light-ring, one may seek Pad\'e approximants which robustly fit the functions across both weak-field and strong-field domains. These approximants may serve to constrain free functions within EOB theories.

As discussed in Ref.~\cite{Bini:Damour:2014}, gauge-invariant kinematical quantities can also have a \emph{dynamical} significance in EOB theory. For example, the `electric-quadrupole' $\EE^2$ features in the leading-order tidal correction to the effective action of the binary system. In Sec.~\ref{subsec:EOB} we showed that our new results can be used to move beyond the 2PN expansion for $\EE^2$ derived in Ref.~\cite{Bini:2012gu}. There are surely more connections of this kind yet to be explored.

We believe that there are no further independent invariants associated with equatorial circular orbits,  if attention is restricted to second derivatives of the regular metric. However, there are certainly `octupolar' quantities, featuring third derivatives, which are also of relevance in EOB theory. We hope our approach will soon be extended to compute such octupolar invariants.

Another challenge for the near future is to compute the spin precession and tidal invariants on the Kerr spacetime. The relevant expressions to be implemented are given in Sec.~\ref{subsec:perturbationtheory}. We hope that the radiation-gauge formalism developed by Friedman, Shah and collaborators \cite{Shah:Friedman:Keidl:2012} may be extended to compute highly-accurate results in the Kerr case (see e.g.~\cite{Isoyama:2014mja} for recent progress).

There is also the prospect of generalizing our approach to encompass non-circular and non-equatorial  trajectories. In more general cases, we anticipate that there will be additional degrees of freedom, with a naive counting suggesting the existence of (up to) three precession quantities, and (up to) seven tidal quantities (cf.~one and four, respectively, for the circular, equatorial case). As these quantities vary around the orbit, it is not immediately clear whether they have a gauge-invariant local meaning, or whether they may only be defined via orbital averages, as in Ref.~\cite{Barack:Sago:2011}.

Another intriguing avenue for future work is the calculation of tidal invariants at second order in the mass ratio (or for general mass ratios). Here, the key point underlying our approach has been that, at $O(\mu)$, the motion of the small body is mapped onto a trajectory in a regular perturbed metric. This intuitively-appealing idea was put on a firm footing by Detweiler \& Whiting \cite{Detweiler:Whiting:2003} and others \cite{Harte:2012}. It seems plausible that a similar interpretation may be possible at higher orders (e.g.~$O(\mu^2)$). Formulations of the second-order problem by Pound \cite{Pound:2012}, Gralla \cite{Gralla:2012}, and Detweiler \cite{Detweiler:2012} have laid a foundation. Recent progress in overcoming certain practical and technical barriers \cite{Warburton:Wardell:2013, Pound:Miller:2014} suggests that second-order results are imminent. Attention will initially focus on the redshift invariant \cite{Pound:2014}, but we hope that calculations of other invariants will follow.

\section*{Acknowledgements}
The authors are grateful to Thibault Damour and Donato Bini for their helpful correspondence regarding the numerical error in the original version of Table I, which is now corrected. SRD thanks Abraham Harte, Adam Pound, Alexandre Le Tiec, Marc Casals and Leor Barack for discussions and guidance. BW thanks David Nichols, Leo Stein and Peter Taylor for helpful conversations. PN and ACO acknowledge support from
Science Foundation Ireland under Grant No.~10/RFP/PHY2847. NW's work was supported by the Irish
Research Council, which is funded under the National Development Plan for
Ireland. BW gratefully acknowledges support from the John Templeton Foundation
New Frontiers Program under Grant No.~37426 (University of Chicago) -
FP050136-B (Cornell University).
\newpage
\appendix

\section{Numerical data for $\Delta U$ and $\Delta\psi$}\label{apdx}

As well as computing the tidal invariants we have used our Regge-Wheeler code to calculate Detweiler's redshift invariant \cite{Detweiler:2008} and the spin invariant \cite{Dolan:etal:2014}. Our results are presented in Table \ref{table:results_U_psi} below.

\begin{table}[H]
\begin{center}
\begin{tabular}{c | c  c }
\hline\hline
$r_\Omega/M$ 	& $\Delta U \times M/\mu$							& $\Delta\psi\times M/\mu$	\\
\hline
$4$	 & $-1.218697151453$	 & $-1.1669040564\times 10^{-1}$	\\
$5$	 & $-4.666523741995578\times 10^{-1}$	 & $-1.6054964918747\times 10^{-2}$	\\
$6$	 & $-2.9602750929001455\times 10^{-1}$	 & $1.8780999340845\times 10^{-3}$	\\
$7$	 & $-2.20847527432247320\times 10^{-1}$	 & $6.09233649269254\times 10^{-3}$	\\
$8$	 & $-1.77719743553592433\times 10^{-1}$	 & $6.81782901966735\times 10^{-3}$	\\
$9$	 & $-1.49360608917907227\times 10^{-1}$	 & $6.52052387967319\times 10^{-3}$	\\
$10$	 & $-1.29122274392049459\times 10^{-1}$	 & $5.93856587591750\times 10^{-3}$	\\
$12$	 & $-1.01935572386267132\times 10^{-1}$	 & $4.73477731157994\times 10^{-3}$	\\
$14$	 & $-8.43819534095711226\times 10^{-2}$	 & $3.76605173794122\times 10^{-3}$	\\
$16$	 & $-7.20550574293450112\times 10^{-2}$	 & $3.03671433760862\times 10^{-3}$	\\
$18$	 & $-6.29018994282390090\times 10^{-2}$	 & $2.48873365079803\times 10^{-3}$	\\
$20$	 & $-5.58277186024938513\times 10^{-2}$	 & $2.07150084940121\times 10^{-3}$	\\
$30$	 & $-3.57783135718205099\times 10^{-2}$	 & $9.90033223034276\times 10^{-4}$	\\
$40$	 & $-2.63396774137048419\times 10^{-2}$	 & $5.75052338252045\times 10^{-4}$	\\
$50$	 & $-2.08446565305954225\times 10^{-2}$	 & $3.74759200441582\times 10^{-4}$	\\
$60$	 & $-1.72475932926791548\times 10^{-2}$	 & $2.63295728928835\times 10^{-4}$	\\
$70$	 & $-1.47096463617217204\times 10^{-2}$	 & $1.95016967400540\times 10^{-4}$	\\
$80$	 & $-1.28229605757714959\times 10^{-2}$	 & $1.50204802830339\times 10^{-4}$	\\
$90$	 & $-1.13653156074114270\times 10^{-2}$	 & $1.19225904925310\times 10^{-4}$	\\
$100$	 & $-1.02052827300276055\times 10^{-2}$	 & $9.69242890897005\times 10^{-5}$	\\
$500$	 & $-2.00804044413976405\times 10^{-3}$	 & $3.97588018220824\times 10^{-6}$	\\
$1000$	 & $-1.00200502771414297\times 10^{-3}$	 & $9.96992511214102\times 10^{-7}$	\\
$5000$	 & $-2.00080040044302370\times 10^{-4}$	 & $3.99759880077002\times 10^{-8}$	\\
\hline \hline
\end{tabular}
\end{center}
\caption{Numerical results for $\Delta U$ and $\Delta\psi$, the redshift and spin precession invariants, respectively. We believe that all the digits presented are accurate.}\label{table:results_U_psi}
\end{table}

\section*{References}

\bibliographystyle{unsrt}


\end{document}